\documentclass[final]{ias2}

\usepackage{graphicx} 
\usepackage{multirow}
\usepackage{array} 

\usepackage{hyperref} 
\usepackage{enumerate}
\begin{document}

\markboth{Pramana class file for \LaTeX 2e}{Ranjini Bandyopadhyay}

\title{Novel Experimentally Observed Phenomena in Soft Matter}

\author[ran]{Ranjini Bandyopadhyay} 
\email{ranjini@rri.res.in}
\address[ran]{Raman Research Institute, CV Raman Avenue, Sadashivanagar, Bangalore, 560080, INDIA}

\begin{abstract}
Soft materials such as colloidal suspensions, polymer solutions and liquid crystals are constituted by mesoscopic entities held together by weak forces. Their mechanical moduli are several orders of magnitude lower than those of atomic solids. The application of small to moderate stresses to these materials results in the disruption of their microstructures. The resulting flow is non-Newtonian and is characterised by features such as shear rate-dependent viscosities and non-zero normal stresses. This article begins with an introduction to some unusual flow properties displayed by soft matter. Experiments that report a spectrum of novel phenomena exhibited by these materials, such as turbulent drag reduction, elastic turbulence, the formation of shear bands and the existence of rheological chaos, flow-induced birefringence and the unusual rheology of soft glassy materials, are reviewed. The focus then shifts to observations of the liquid-like response of granular media that have been subjected to external forces. The article concludes with examples of the patterns that emerge when certain soft materials are vibrated, or when they are displaced with Newtonian fluids of lower viscosities. 
\end{abstract} 


 
\maketitle

% \tableofcontents
% \listoffigures
% \listoftables

\section{Introduction}

\subsection{Soft Materials}

\hspace{0.25cm} The beginning of the last century saw the development of new synthetic materials such as polymers and paint that display seemingly odd flow behaviours. The realisation that their responses to external stresses cannot be described by Hooke's law of elasticity \cite{hooke_elasticity} or Newton's law of viscosity \cite{newton_viscosity} led Eugene Bingham to coin the term rheology (derived from the Greek verb $\rho\epsilon\iota\nu$ meaning 'to flow')  for the study of their flow and deformation \cite{markovitz_rheo,macosko_rheo,barnes_rheo,mezger_rheo,dorai_rheo}. Shaving foams, shampoo, blood, ink, ketchup and soap solutions are some everyday examples of non-Hookean and non-Newtonian materials  \cite{piazza_soft,mitov_soft}. The building blocks of these materials are supramolecular aggregates. The forces  that exist between these aggregates (typically, screened Coulomb repulsions, hydrogen bonds and van der Waals attractions) can be overcome by small or moderate stresses.  They are therefore called soft materials. Besides having a wide range of applications, soft materials such as liquid crystals, aqueous colloidal suspensions, polymer solutions and granular materials are also important for fundamental research \cite{hamley_soft}. 

French physicist Pierre-Gilles de Gennes, often described as the founding father of soft matter physics, began his Nobel lecture with the following sentences \cite{degennes_nobel}: {\it `What do we mean by soft matter? Americans prefer to call it “complex fluids”. This is a rather ugly name which tends to discourage the young students. But it does indeed bring in two of the major features'}. de Gennes went on to list {\it complexity} and {\it flexibility} as these two major features. 

The {\it complexity} of soft matter arises from the elaborate organisation of the supramolecular building blocks. de Gennes illustrated the {\it flexibility} of soft matter by citing the  example of the boots that Amerindians of the Amazon basin made to protect their feet  \cite{degennes_nobel}. They extracted latex  (a viscous solution of unentangled, long-chain polymers) from  Hevea trees, which they applied to their feet. The latex solidified quickly into something resembling a boot. This liquid-solid transition is brought about by the oxygen in the atmosphere that reacts with latex to form bridges between the polymeric chains at specific points. While unconnected polymer chains exhibit viscous flow, the polymer network that is formed when latex reacts with oxygen is elastic. 

There is, however, more to the story of the boots. These boots are rather short-lived as the oxygen, which initially bound the polymer chains, eventually ends up severing those very bonds that it helped make. In 1839, Charles Goodyear \cite{goodyear_patent} showed that if latex is boiled with suphur (in a process called vulcanisation), the rubber that is formed is extremely strong and resilient \cite{degennes_fragile}. The  Amerindians' boots and the vulcanised rubber both reveal the enormous flexibility of latex and are excellent examples of how a small chemical reaction is sufficient to completely alter its flow behaviour. 

This article focusses on the flexibility of soft matter. Its aim is to introduce the reader to the intriguing dynamics of soft matter that are triggered, not by chemical reactions, but by the application of small or moderate stresses and strains.

\subsection{Elasticity, viscosity and viscoelasticity}

\hspace{0.25cm}The behaviour of solids subjected to external forces is described by Hooke's law: $\sigma$ = G${\gamma}$ \cite{hooke_elasticity}. Here, $\sigma$ is the stress, or the force per unit area,  and $\gamma$ is the shear strain, or the relative change in length. G is the elastic modulus and is an intrinsic property of the deformed material.  Most metals and ceramics show Hookean behaviour when subjected to low strains \cite{macosko_rheo}. Liquids such as water and some oils, on the other hand, are described by Newton's law of viscosity: $\sigma$ = $\eta{\dot\gamma}$ \cite{newton_viscosity} (Fig.\ref{fig:rheo}(a)). Here, $\dot\gamma$ is the rate at which the applied strain changes. The proportionality constant $\eta$, another material constant, is the viscosity of the liquid and measures its lack of 'slipperiness' or its resistance to flow when an external shear stress is imposed. 

Soft materials are neither perfectly elastic nor perfectly viscous. Instead, they exhibit both elastic and viscous responses and are called `viscoelastic materials' or `non-Newtonian fluids'. Their flow is very sensitive to their structural organisation, with their microstructures being easily altered by external stresses. When a Hookean solid is deformed, the extension is  instantaneous. In contrast, the stress generated by a deformed viscoelastic material is a nonlinear function of the history of the deformation gradient. Viscoelastic materials are therefore  also referred to as memory fluids. If a rotating shaft that is stirring a polymeric liquid is suddenly released, the shaft will return halfway. However, if the shaft is held stationary for some time before its release, the extent of reformation is considerably smaller because of the fading memory of the polymeric liquid.  Clearly,  Hooke's law and Newton's law of viscosity are inadequate to describe this experiment. For a video of this experiment and also of several other demonstrations of the intriguing flow properties of soft materials, the reader is referred to \cite{ncfm_rheologynon}.

\subsection{The stress tensor} 

\hspace{0.25cm}The stress tensor \cite{landau_elasticity,feinman_vol2} provides a complete description of the stresses within a three dimensional object. The scalar component $\sigma_{ij}$ of the stress tensor, representing the component of the stress in the $i$ direction on a surface whose normal is in the $j$ direction,  is written as
\begin{equation}
\sigma_{ij} = \begin{bmatrix} \sigma_{xx} &  \sigma_{xy} &  \sigma_{xz} \\ \sigma_{yx} &  \sigma_{yy} &  \sigma_{yz} \\\sigma_{zx} &  \sigma_{zy} &  \sigma_{zz}\end{bmatrix}.
\end{equation}
  
A Newtonian liquid at rest supports only a uniform normal stress, its hydrostatic pressure $p$ ($\sigma_{xx}$  =  $\sigma_{yy}$  =  $\sigma_{zz}$ = -$p$).  When this liquid is sheared, the Newtonian constitutive relation \cite{newton_viscosity} gives $\sigma_{xy}$ = $\sigma_{yx}$ = $\eta\dot\gamma$. In viscoelastic materials, $\sigma_{xx} \neq \sigma_{yy} \neq \sigma_{zz}$ and the first and second normal stress differences ( N$_{1}$ = $\sigma_{xx}$ - $\sigma_{yy}$ and  N$_{2}$ = $\sigma_{yy}$ - $\sigma_{zz}$)  are non-zero. For large deformations, the shear stress $\sigma_{xy}$ is a function of the deformation field. These features give rise to novel and counter-intuitive flow phenomena and will be discussed in the subsequent sections. 

\section{The rheology of soft materials}

\begin{figure}[!t]
\begin{center}
\includegraphics[width=1.1\columnwidth]{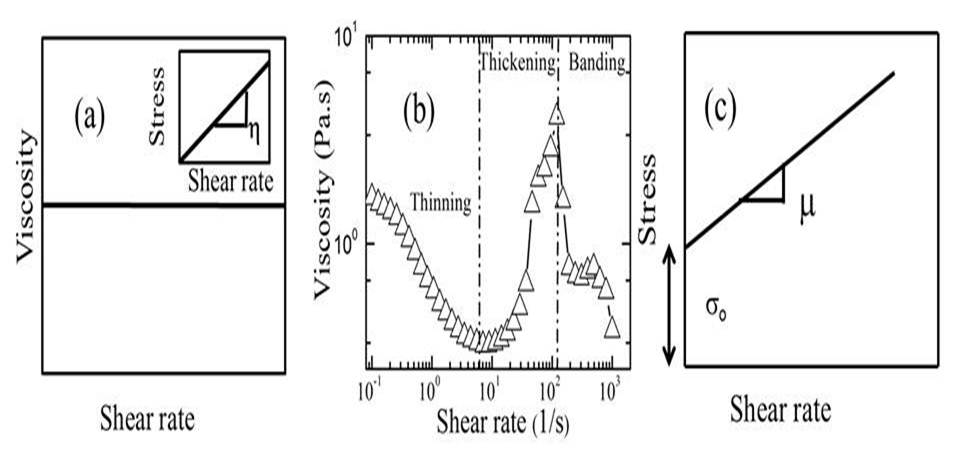}
\caption{ (a) In a Newtonian liquid, the viscosity is independent of the applied shear rate. Inset: The stress varies linearly with shear rate and the slope of this plot gives the viscosity $\eta$ (b) The viscosity of a 35 wt.\% aqueous cornstarch suspension changes dramatically with changes in shear rate \cite{rajib}. As shear rate is increased, the suspension viscosity first decreases (shear-thinning) and then increases (shear-thickening). Finally, shear-banding takes over and the suspension breaks up into coexisiting bands of high and low viscosities. (c) The schematic flow curve of a Bingham plastic is described by its yield stress $\sigma_{\circ}$ and plastic viscosity $\mu$. }
\label{fig:rheo}
\end{center}
\end{figure}

\hspace{0.25cm}Wilhelm Weber noticed in 1835 that loading a silk thread resulted in an initial instantaneous extension that was followed by a more gradual increase in the length of the thread. When the load was removed, the thread eventually contracted to its original length \cite{weber_silk}. Such a time-dependent response is a typical feature of viscoelastic materials and can be described by a time-dependent modulus G(t) = $\sigma_{xy}$(t)$/\gamma$ \cite{macosko_rheo}. This relation holds in the limit of small strains, which is called the linear viscoelastic regime (LVE). For large applied strains, the shear modulus also depends on the  strain. The resultant `nonlinear' modulus is defined as G$_{nl}$(t,$\gamma$) = $\sigma_{xy}(t,\gamma)$/$\gamma$ \cite{macosko_rheo}. 
\begin{figure}[!t]
\begin{center}
\includegraphics[width=1.0\columnwidth]{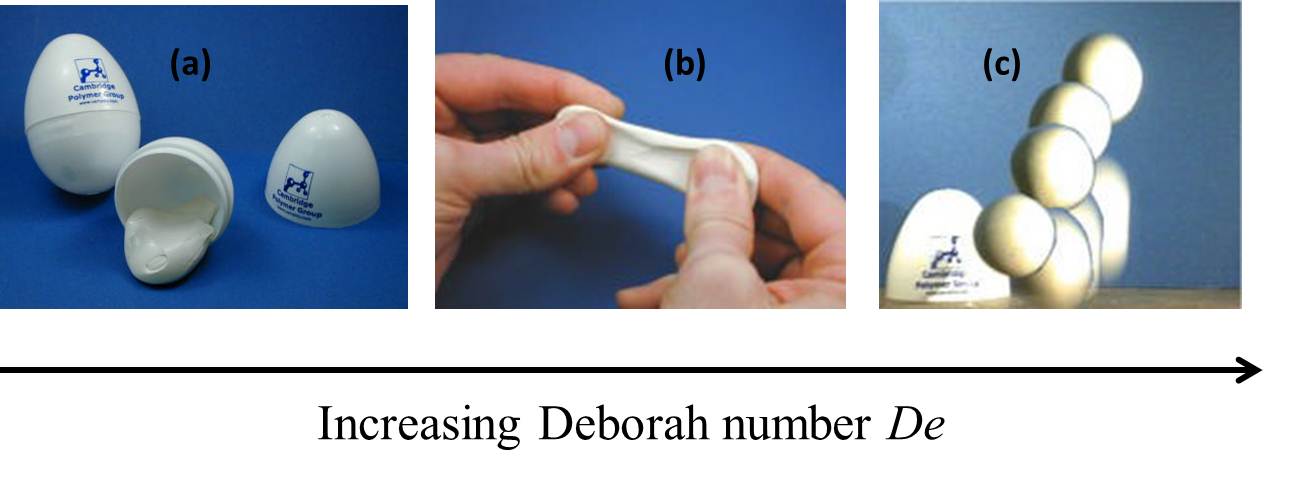}
\caption{The responses of  Silly Putty$^{TM}$ with changing Deborah number $De$. (a) At small $De$, Silly Putty$^{TM}$ flows due to the action of gravitational stresses. (b) At intermediate $De$, it stretches without breaking. (c) At large $De$,  it bounces like a ball and exhibits solid-like behaviour. This figure is reproduced with permission \cite{pol_sillputt}.}
\label{fig:sillyputty}
\end{center}
\end{figure}

Linear viscoelastic flow can be modelled by linear combinations of springs (perfectly elastic elements) and dashpots (perfectly dissipative elements). The simplest models are the Kelvin-Voigt \cite{voigt_linear} and the Maxwell \cite{maxwell_linear} models and comprise one spring and one dashpot, in parallel and series, respectively. Examples of non-Newtonian flow are displayed in Figs. 1(b) and 1(c) respectively \cite{macosko_rheo,laurent_petri, rajib}. 

\subsection{Dilatancy and shear-thickening}

\hspace{0.25cm}If the viscosity of a material increases with the rate of deformation, it is a dilatant or a shear-thickening material \cite{macosko_rheo}. The viscosity of a wide range of viscoelastic materials can be modelled by the relation $\eta = K\dot\gamma^{n-1}$, where $\dot\gamma$ is the applied shear rate, and $n$, $K$  are material constants \cite{barnes_rheo,macosko_rheo}. These materials, called `power-law fluids',  are usually shear-thinning (with $n <$ 1, as in polymer melts). In contrast, some materials, such as concentrated suspensions, show regimes of shear-thickening (with $n >$ 1). A good example of a material that can exhibit both shear-thinning and shear-thickening is a concentrated aqueous suspension of cornstarch (a fine, powdery starch, extracted from maize or corn) which `jams'  when it is stirred vigorously, but flows when stirred gently  (\cite{youtube_cornstarch1,fall_prl}, Fig.\ref{fig:rheo}(b)). The jamming phenomenon is triggered by the shear-driven crowding of the macromolecules, which imposes kinetic constraints on the macromolecular dynamics, thereby increasing the viscosity of the suspension \cite{liu_jam}. When a projectile impacts a cornstarch suspension, large, positive normal stresses develop, and shear-thickening alone cannot explain the observed solidification. A recent work shows that when concentrated suspensions of cornastarch are compressed, dynamic jamming fronts that can absorb a large amount of momentum develop, which results in the growth of solid-like columns in the suspension \cite{jaeger_nature}. 

Strain-induced slowdown of the dynamics of jammed aqueous foams and aging clay suspensions have  been reported in \cite{gopal_prl,band_SM}. Dynamical slowing down leads to shear-thickening in both these materials can be attributed to jamming, which results in the confinement of the foam bubbles and the electrostatically charged clay platelets. 
A commonly cited example of dilatancy is the `wet sand effect'. When a person walks on compacted wet sand on a beach, the sand swells (dilates) around his or her feet \cite{reynolds_dilatancy}. The water trickles down through the pores, leaving dry sand above. 

 Silly Putty$^{TM}$ is an easily available shear-thickening material \cite{crayola}. It is marketed widely as a toy and is composed of silicone polymers such as polydimethylsiloxane. As illustrated in Fig. \ref{fig:sillyputty}, Silly Putty$^{TM}$ can bounce like an elastic solid at short times, but can spread or flow like a viscous liquid at long times. When pulled slowly, it behaves like a liquid and stretches without breaking. For sudden pulls, it snaps like a solid. A useful dimensionless number to introduce here is the Deborah number $De$ = $\tau/\tau_{flow}$, where $\tau$ is the characteristic relaxation time of the material and $\tau_{flow}$ is the time scale of the externally imposed flow. When $De$ is large ($\tau_{flow} <<$ $\tau$), Silly Putty$^{TM}$ retains memory of its size and shape and behaves like an elastic solid. If $De$ is small ($\tau_{flow} >> \tau$), it flows like a liquid due to its fading memory.  While the viscosity of a shear-thickening material increases with an increase in the rate of shear, the viscosity of a rheopectic materials, such as printers ink or gypsum, increases with time due to the application of a constant shear stress. 

\subsection{Shear-thinning}

\hspace{0.25cm}In 1923, F. Schalek and A. Szegvai showed that aqueous iron oxide gels, when shaken, liquefied to exhibit sol-like properties  \cite{szevgai,schalek_thixo}. The sample solidified again when the shaking was stopped. The term 'thixotropy', derived from the Greek words `thixis' (to touch) and `trepo' (to change), describes the reversible behaviour of materials that are solid-like under normal circumstance, but that flow when sheared or stirred. Their viscosities depend on the time duration, and not the rate, of the applied shear. Thixotropic materials do not attain steady-state flow instaneously when a constant shear stress is applied. They also typically require some time to return to a steady state after the cessation of shear. An excellent review on this aspect of non-Newtonian flow can be found in \cite{rheol_cambridge}. Paints and adhesives are two common materials that exhibit thixotropic flow. 

\begin{figure}[!t]
\begin{center}
\includegraphics[width=0.9\columnwidth]{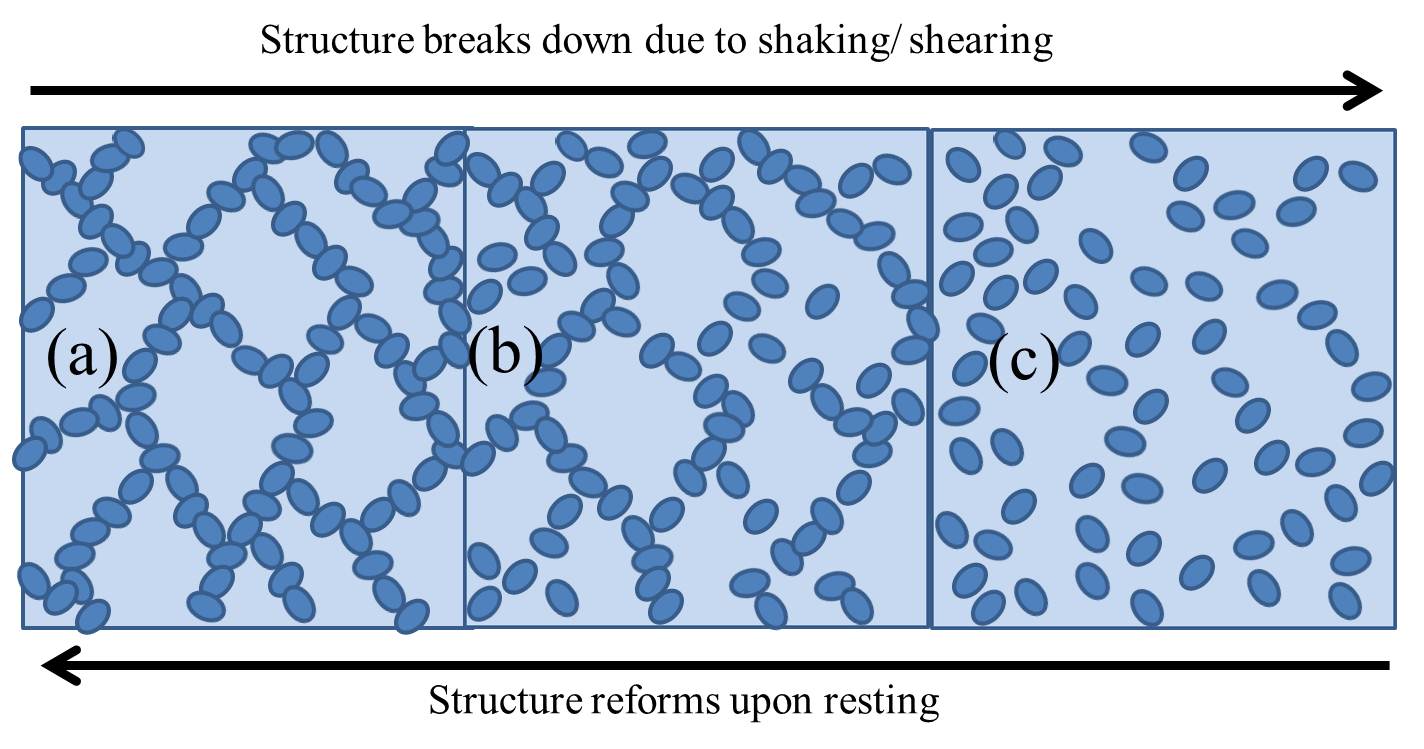}
\caption{The structure of a thixotropic material can be broken down by external shears and can reform spontaneously when the shear is withdrawn. (a) When a very small shear is applied, the network structure of a gel is retained and the sample responds like an elastic solid. (b) Partial structural breakdown occurs with increase in shearing and results in a viscoelastic reponse. (c) When even higher shears are applied, the structural breakdown is complete and shear-thinning flow is seen. This figure is adapted from \cite{barnes_thixo}. }
\label{fig:thixo}
\end{center}
\end{figure}

Shear-thinning or pseudoplasticity refers to the decrease in a material's viscosity with increasing rate of shear (Fig. \ref{fig:rheo}(b), Fig. \ref{fig:thixo}, \cite{barnes_thixo}). All thixotropic materials are therefore shear-thinning, although  the converse is not necessarily true. When a highly viscous solution of entangled polymers flows in a pipe, shear forces disentangle the polymer chains. If the shear forces are high enough, the polymer chains eventually align along the flow direction. This results in a sharp drop in the viscosity of the sample \cite{macosko_rheo,ferry_polymer}. Melts of the polymer acrylonitrile butadiene styrene (ABS) show power-law shear thinning \cite{cox_thin}. For low $\dot\gamma$, the viscosity has a constant magnitude $\eta_{\circ}$ (the zero shear viscosity). For intermediate $\dot\gamma$, $\eta(\dot\gamma) = m{\dot\gamma}^{n-1}$, where $m$ and $n  (<1)$ are constants.  At higher $\dot\gamma$, Newtonian behaviour is sometimes recovered. 

Aqueous foam, which is constituted by closely-packed polydisperse gas bubbles dispersed in a small volume of surfactant solution, can exhibit shear-thinning. Shaving foam, when sprayed on a surface, retains its shape and behaves like a solid, but can flow by the stick-slip motion of the bubbles when gently tapped with a finger \cite{weitz_foam}. Similarly,  marmalade on a piece of toast retains its shape when left unperturbed, but flows easily under the shearing action of a knife. Suspensions, which are mixtures of two substances, one of which is finely divided and dispersed in the other (for example, colloidal particles in water), and emulsions  which are tiny droplets of one material suspended in another immiscible material (for example, milk, mayonnaise and paint) exhibit power-law shear-thinning due to flow-induced microstructural rearrangements \cite{barnes_thixo}.  

An interesting consequence of shear-thinning is the Kaye effect \cite{kaye_thin}. When a stream of shampoo hits a surface, a heap is created at the point of impact. The shampoo shear-thins as it flows down the heap under gravity. As more shampoo falls on this shear-thinned layer, it slips and occasionally flies off a dimple in the underlying heap. This results in secondary lasso-shaped jets of shampoo that hit the primary stream of shampoo before finally collapsing. This is called the Kaye effect and usually lasts less than a few hundred milliseconds \cite{laurent_petri,versluis_kaye,versluis_youtube}. 

\subsection{Yielding and plastic flow}

\hspace{0.25cm}A material exhibits plastic behaviour if it does not flow when a stress lying below a certain threshold (called the yield stress) is applied. If it flows like a Newtonian liquid when the yield stress is exceeded, it is called a Bingham plastic \cite{bingham_yield}. Such a flow can be described as follows: $\dot\gamma$ = 0 for  $\sigma < \sigma_{o}$, and $\sigma - \sigma_{\circ} = \mu{\dot\gamma}$ for $\sigma \ge \sigma_{o}$ (Fig.\ref{fig:rheo}(c)).  Here, $\sigma_{o}$, the yield stress \cite{macosko_rheo}, is a measure of the stress required to break the  structure of the sample. 

Toothpaste, mayonnaise, tomato ketchup and aqueous foam are examples of materials with  non-zero yield stresses. Toothpaste, for instance, cannot flow out of its tube spontaneously, but oozes out when the tube is squeezed hard enough that the yield stress is exceeded. Similarly, tomato ketchup has a non-zero yield stress and flows out of the bottle only after vigorous tapping. When subjected to stresses that exceed the yielding value, emulsions flow due to the cooperative motion of the rearranging droplets \cite{goyon_yield, boquet_yield}. The extent of the spatial cooperativity can be described in terms of a correlation length that diverges close to the yielding threshold.  The solid-liquid transition in normal emulsions is continuous, and the static (solid-liquid) and the dynamic (liquid-solid) yield stresses are identical.  When thixotropic emulsions (composed of clay platelets forming links between the emulsion droplets) are sheared, the transition from rest (solid) to flow (liquid)  is discontinuous. Localised shear zones (shear bands) exist in the no-flow regime and flow occurs only above a sample history-dependent shear rate $\dot\gamma_{c}$ \cite{fall_yield}. For normal emulsions, $\dot\gamma_{c} \rightarrow$ 0.

The viscosity of an ideal yield stress fluid is expected to diverge continuously as the yield stress is approached. The viscosity of a real yield stress fluid such as a clay suspension, in contrast, diverges abruptly at the yield point \cite{coussot_yield}. The second observation is explained in terms of a bifurcation in the rheology of the clay suspension. For very low stresses, the clay sample ages, its viscosity increases with time  and the flow soon stops. For large applied stresses, the sample shear-thins (in a phenomenon called `rejuvenation') and  the sample flows like a liquid. Very close to the yielding threshold, the flow is characterised by the presence of avalanches. 

%\begin{figure}[!t]
%\begin{center}
%\includegraphics[width=0.8\columnwidth]{yield.jpg} 
%\caption{Mayonnaise, a material with a non-zero yield stress is shown in (a). It behaves like a solid at rest or when subjected to %stresses below the yielding value. This is evident from the peaks and ridges on its surface. This image is reproduced from %\cite{wiki_bingham} under a Creative Commons License. The schematic flow curve of a Bingham plastic having yield stress %$\sigma_{\circ}$ and plastic viscosity $\mu$ is shown in (b). }
%\label{fig:yield}
%\end{center}
%\end{figure} 

\subsection{Non-zero normal stresses and extensional thickening} 

\hspace{0.25cm}The non-linearities inherent in the stress-deformation relation of sheared viscoelastic materials gives rise to non-zero normal stress differences. When a viscoelastic material is subjected to moderately large values of $\dot\gamma$, the first and second normal stress differences  are given by $N_{1}$ = $\sigma_{xx}$ - $\sigma_{yy}$ = -$\psi_{1}\dot\gamma^{2}$ and  $N_{2}$ = $\sigma_{yy}$ - $\sigma_{zz}$ = -$\psi_{2}\dot\gamma^{2}$. Here,  $\psi_{1}$ and $\psi_{2}$,  the first and second normal stress coefficients, are positive and negative, respectively \cite{macosko_rheo}.

The viscosities of melts of ABS and PS (polystyrene) decrease with increasing rate of shear deformation, but  increase in elongational flow \cite{cox_thin}. This, phenomenon, called extensional thickening, arises due to the non-zero value of $N_{1}$. The corresponding time-dependent uniaxial extensional viscosity, $\eta^{+}_{u}(t,\dot\gamma)$ = $[\sigma_{xx}(t,\dot\gamma) - \sigma_{yy}(t,\dot\gamma)]/\dot\gamma$, cannot be inferred from the shear viscosity $\eta$.  A small change in the composition of polymer solutions can cause an appreciable change in $\eta_{u}^{+}$ and almost no change in $\eta$ \cite{chao_pol, bird_rheo}. In the low $\dot\gamma$ regime, however, the prediction of Trouton's law ($\eta_{u}^{+}$ = 3$\eta_{\circ}$) works quite well \cite{trouton_extvis}.

\section{Some novel consequences of non-Newtonian flow}

Due to their shear rate-dependent viscosities, the existence of yield stresses and non-zero normal stresses, soft materials exhibit novel,  nonintuitive dynamics, a select few of which will be highlighted in this section.

\subsection {The unusual flow of polymer solutions}
If a rotating disk is placed on the surface of a beaker filled with a Newtonian liquid, the liquid is pushed outward towards the walls of the beaker. It then descends along the walls and eventually rises at the centre. When a similar experiment is repeated with a beaker containing a polymer solution, the currents are in the opposite directions. In another experiment, a disk rotating at the bottom of a beaker filled with a Newtonian liquid causes a depression in the liquid level at the surface. A disk rotating at the bottom of a beaker filled with a polymeric solution, in contrast, causes the solution surface to rise at the centre \cite{ptoday_bird}.  These observations arise due to the positive value of the first normal stress coefficient $\psi_{1}$. It is well-known that the surface of a Newtonian liquid flowing down a tilted trough of semi-circular cross section is flat, apart from some curvature due to meniscus effects. A non-Newtonian liquid flowing down the same trough, however, has a convex surface. This is a consequence of the negative value of its second normal stress coefficient $\psi_{2}$ \cite{ptoday_bird}. The remainder of this subsection discusses three examples of the unusual flow of polymer solutions. The reader is refered to \cite{ptoday_bird,khan_rheo,mit_demo} for detailed discussions of these and related phenomena. 

\begin{figure}[!t]
\begin{center}
\includegraphics[width=1.1\columnwidth]{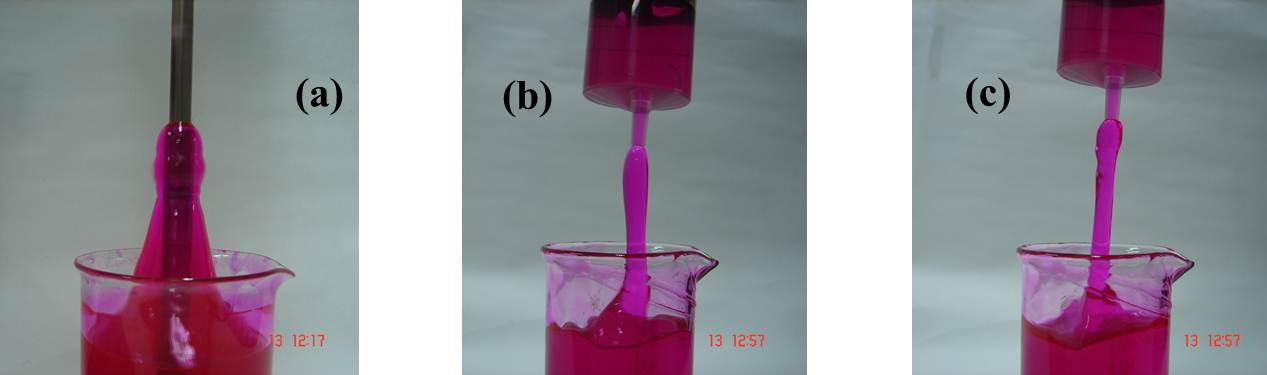} 
\caption{ (a) {\it Weissenberg effect:} when an elastic polymer solution kept in a beaker is stirred, the solution climbs the stirring rod. (b)  {\it Barus effect:} the polymer solution swells when squeezed through an orifice. (c)  {\it Fano flow:} the polymer solution can be sucked out of a beaker with a syringe that is not in contact with the surface of the sample. All the experiments are performed with a dyed, 2 wt.\% polyacrylamide (PAM) solution \cite{yogesh_lab}. Videos of these experiments have been uploaded as supporting information \cite{yogesh_videos}.}
\label{fig:pol}
\end{center}
\end{figure} 
\begin{enumerate}[(a)]
\item{\it Weissenberg effect:}
 When a  Newtonian liquid is stirred with a rod, the inertial forces that are generated throw the liquid outward. This causes a depression in the liquid level near the rod and an elevation at the walls. In contrast,when a polymer solution is stirred, it climbs the rotating rod. This phenomenon, also referrred to as the rod-climbing effect \cite{Weissenberg,kundu_weissenberg,mit_weissenberg}, is illustrated in Fig. \ref{fig:pol}(a). When the viscoelastic polymer solution is stirred, a tension develops along the circular lines of flow. This results in a pressure that is directed towards the centre of the beaker. A finite, positive $N_{1}$ develops, which drives the polymer solution up the rotating rod.

%\begin{figure}[!t]
%\begin{center}
%\includegraphics[width=0.5\columnwidth]{dieswell.jpg} 
%\caption{Schematics of  a (a) Newtonian fluid and a (b) polymeric fluid being pushed out through an orifice. In (a), the diameter %of the Newtonian fluid is the same as the orifice size immediately after exiting the orifice. In contrast, there is a significant %swelling in the radius of the polymeric fluid immediately after exiting the orifice and can be seen in (b).}
%\label{fig:die}
%\end{center}
%\end{figure} 

\item{\it Barus effect:} When a polymer solution is extruded through an orifice, it swells (Fig. \ref{fig:pol}(b)). The swelling can be upto three or four times of the orifice diameter. This  phenomenon is also referred to as the Merrington effect, die swell or extrudate swell. As the sample is squeezed out through the orifice, the stretching of the polymer chains results in a non-zero value of $N_{1}$. Immediately after leaving the orifice, the polymer chains relax back to their original globular structures. This gives rise to the observed swelling. As $N_{1}$ increases with shear rate $\dot\gamma$, extrudate swelling increases with increasing rate of extrusion  \cite{tanner_dieswell,ptoday_bird,youtube_barus,mit_demo}. This example demonstrates that the polymer chains retain memory of their deformation history during the extrusion process. Delayed die swell of elongated macromolecules has been reported in wormlike micellar solutions \cite{cloitre_JNNFM}.

\item{\it Fano flow:} %\begin{figure}[!b]
%\begin{center}
%\includegraphics[width=0.56\columnwidth]{fano.jpg} 
%\caption{A schematic diagram showing that (a) a Newtonian fluid can only be siphoned out when the syringe is dipped under the %fluid. In sharp contrast, a polymeric fluid can be siphoned out even when the syringe is at a considerable height above the fluid %surface and is shown in (b). }
%\label{fig:fano}
%\end{center}
%\end{figure}
This is a common example of extensional thickening \cite{fano_ext}. If a nozzle dipped into a bath containing a Newtonian liquid is raised above the liquid surface, the liquid column breaks immediately. To siphon out  a Newtonian liquid, the nozzle must be dipped below the surface. Non-Newtonian fluids, in contrast, can be pulled up to a considerable height above the fluid surface without rupturing the fluid column \cite{macosko_rheo,mit_demo}. This phenomenon, also referred to as a `ductless' or a `tubeless' siphon, is illustrated in Fig. \ref{fig:pol}(c). When a highly viscoelastic material is sucked out, large normal stresses that balance the weight of the fluid column develop. This results in Fano flow \cite{tumblr_fano}. 

\end{enumerate}
\subsection {Turbulent drag reduction:} 
 
\hspace{0.25cm}In 1948, B. Toms noticed that if a very minute quantity (ten parts per million by weight) of the polymer poly(methyl methacrylate)  (PMMA) was added to a Newtonian liquid (monochlorobenzene) flowing through a pipe at high Reynolds numbers, the turbulent motion of the latter could be suppressed considerably \cite{toms_drared}. It is now understood that the addition of PMMA reduces the pressure drop across the pipe and results in a significant reduction in skin friction.  A reduction in turbulent flow is also observed when the polymer polyisobutylene (PIB) is dispersed in benzene or cyclohexane, and when PEO or PAM are dispersed in turbulent water \cite{gyr_drared}. When the concentration of polymer additives is gradually increased, drag reduction (DR) increases upto a saturation value, before eventually decreasing \cite{bark_jfm}.  This phenomenon is widely utilized in irrigation networks, sewerage systems, oil pipelines and for extending the range of fire-fighting equipment. Flexible polymers of very high molecular weights, typically composed of  10$^{4}$ - 10$^{5}$  monomeric units, are the most effective as drag reducing agents \cite{fas_dragred}. 

For a closed turbulent flow with zero mean velocity, two types of forcings were used in \cite{bon2_drared}: a smooth forcing in which the fluid motion was driven by the viscous boundary layers, and a rough forcing in which the fluid was stirred with baffles. When a minute quantity of a polymer (thirty parts per million by weight) is added, there is a significant reduction in the turbulent energy dissipation in the smooth forcing experiment. In the rough stirring experiment, no such reduction is seen. Although the understanding of this phenomenon is still far from complete, DR is generally believed to originate from boundary layer effects \cite{bon2_drared}. It should be noted here that some experiments reported in the literature imply the existence of very different mechanisms driving DR. For example, another study has shown that when the polymer is injected at the centre of the pipe, fluid turbulence is reduced even before the injected polymer reaches the boundary walls \cite{frings_drared}. 

The introduction of a polymer additive results in significant changes in the cascade of eddies generated in a Newtonian liquid. This alters its dissipation behaviour \cite{lumley_review}. DR is observed when the wall shear stress is larger than a threshold value, as the flow must first stretch the polymers out from their initial globular configurations \cite{lumley_review}.  It has already been pointed out that while the addition of a small amount of polymer does not change the shear viscosity of a fluid significantly, its elongational viscosity can increase by many orders of magnitude \cite{bird_rheo}.  In \cite{bonn_dragred}, common salt was added to increase the flexibility of polyelectrolyte chains ($\lambda$-DNA and hydrolysed polyacrilamide). This makes the chains more resistant to stretching out in elongational flows and increases the elongational viscosity of the solution. When small amounts of these polyelectrolytes are added to a turbulent Newtonian fluid, DR increases almost linearly with increase in the solution's elongational viscosity.  DR also increases with increase in the Reynolds number $ Re = \frac{\rho{\Omega}Rd}{\eta}$. Here, $\rho$ and $\eta$ are respectively the density and shear viscosity of the fluid, $\Omega$ is the speed of the rotational motion imposed to generate turbulence, and $R$ and $d$ are the radius of the inner cylinder and the gap of the Couette geometry, respectively.  In \cite{wagner_epl}, NaCl was again added to tune the persistence lengths of $\lambda$-DNA chains in an aqueous solution. These experiments demonstrate that turbulent energy dissipation reduces with chain flexibility, and a minimum chain flexibility is essential for DR. In a $\lambda$-DNA solution with no added NaCl, a significant enhancement in drag is observed.  

Surfactants present in water during turbulent pipe flow can reduce drag effectively when they self-assemble into rod-like micelles. The extent of DR is a strong function of the micellar size, number density and surface charge \cite{hoff_drared}. For every micellar system investigated, there exists an absolute temperature $T_{l}$, which depends on the hydrophobic chain length of the surfactant and the concentration of counterions in solution, above which there is no DR. It is now believed that DR in these systems is facilitated by the formation of shear-induced structures that increase wall-slip \cite{drappier_drared, white_drag}.   

\subsection{Elastic turbulence:} 
\hspace{0.25cm}Fluid turbulence occurs when  inertial  driving forces are much larger than viscous damping forces \cite{reynolds_turb} and is accompanied by an enhancement in drag and in spatial and temporal velocity fluctuations. It was shown in \cite{groisman_turb,larson_news} that these features can exist even  in the low Reynolds number flow  ($Re \approx$ 1) of highly elastic polymer solutions. In contrast, the critical Reynolds number for the onset of turbulence in Newtonian liquids, $Re_{crit} \approx$ 10$^{5}$ \cite{landau}. 

Clearly, inertia is negligible in the turbulent flow of polymer solutions. The role of inertia is played by elastic stresses and the flow is  described in terms of an elasticity parameter $Wi/Re = \tau\nu/L^{2}$, where $Wi = {\dot\gamma}{\tau}$ is the Weissenberg number. Here, $\nu$ is the kinematic viscosity and $L$ is a characteristic length scale.  This phenomenon, called elastic turbulence,  occurs at high $Wi$ and low $Re$ and is driven by the nonlinearities that are inherent in the mechanical response of the polymer solution \cite{groisman_turb}.  Polymer molecules, when stretched by flow, become unstable and give rise to irregular secondary flows. These secondary flows, through a feedback mechanism, stretch the polymers even further and fully developed turbulence eventually sets in. Elastic turbulence is characterised by a large increase in the flow resistance and, in sharp contrast to inertial turbulence,  increases with increase in solution viscosity. It was shown in \cite{groisman_turb} that when polymers are suspended in highly viscous sugar solutions, turbulence could be generated at even lower $Re$ ($\approx$ 10$^{-3}$). 

When a melt of linear, low-density PE of viscosity $\eta$ = 5000 Pa.s flows out of a capillary tube above a critical flow rate, extrudate distortions are observed at low $Re$ \cite{kalika_jrheol}. The minimum flow rate for the onset of these distortions decreases when the melt viscosity is increased by increasing the chain length of the polymer.  Elastic vortices were observed in the Taylor-Couette flow of  highly viscoelastic PIB solutions above a critical $Re$ \cite{larson}. For a review on the instabilities observed in viscoelastic flows, the reader is referred to  \cite{larson_review}. 

A complete understanding of these instabilities is still lacking and experiments need to be performed in different geometries and with other non-Newtonian materials \cite{larson_news}. Turbulence and tumbling have been observed  in liquid crystals \cite{manneville_nematic,cladis_lc}. Elastic turbulence and Taylor-like vortices were reported in the Coutte flow of wormlike micellar solutions undergoing shear banding \cite{fardin_turb, fardin_turb2}, and in the curvilinear flow of polymer solutions \cite{groisman_turb2}. It was demonstrated that two very viscous liquids flowing in a curved channel can be mixed very efficiently at low $Re$ if very small quantities of polymers are added to the former \cite{groisman_nature2}.  Elastic turbulence can therefore have very important consequences in the industry.

\subsection{Shear bands and rheological chaos in giant wormlike micellar solutions}

\begin{figure}[!t]
\begin{center}
\includegraphics[width=1.1\columnwidth]{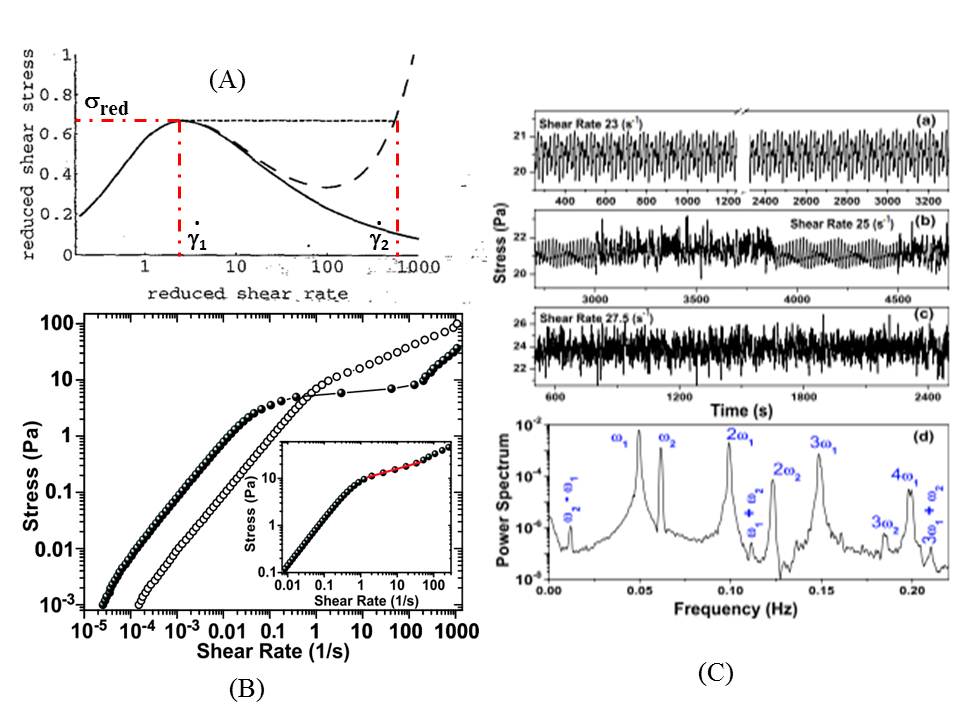} 
\caption{(A): The theoretical flow curve of a sheared GWM solution  is reproduced with permission from \cite{spenley_gwm}. (B)  The flow-concentration coupling in a GWM-salt sample  results in a positive slope in the metastable region of the flow curve and is shown for an aqueous sample of 2 wt.\% CTAT + 100 mM NaCl by hollow circles. Solid circles represent the flow curve of a 2 wt.\% CTAT sample without salt. The inset shows the power-law fit to the metastable region of the flow curve acquired for the sample containing salt. (C) Temporal stress fluctuations for applied shear rates $\dot\gamma$ = (a) 23, (b) 25 and (c) 27.5 /sec are displayed. (d) The Fourier transform of the time-series data acquired at $\dot\gamma$ = 23 /sec shows 2-frequency quasiperiodicity. The data in (B) and (C) are reproduced with permission from \cite{ganapathy_gwm}.}
\label{fig:shear}
\end{center}
\end{figure} 

\hspace{0.25cm}The previous sections demonstrated how polymer solutions and melts, gel networks and colloidal suspensions respond to shear and elongational flows.  In this section, we turn our attention to structure-flow correlations in sheared surfactant aggregates. 

When surfactants, which are amphiphilic macromolecules consisting of a hydrophilic (water-loving) `head' and a hydrophobic (water-hating) `tail', are dissolved in water at certain concentrations and temperatures, they self-assemble to form spherical micelles, long chain-like supramolecular structures called giant wormlike micelles  (GWMs), lamellar phases, onions {\it etc.}  \cite{surfactants}. When aqueous solutions of GWMs are subjected to high shear rates, the wormlike chains disentangle, break, and eventually stretch in the direction of flow \cite{cates_adv}.  A spontaneous retraction accompanies this stretching and gives rise to a stress peak in the flow (stress-{\it vs.}-strain) curves of these samples. Such a  flow curve was first proposed in \cite{spenley_gwm} and is shown in Fig. \ref{fig:shear}(A).

If a shear rate lying in the non-monotonic region of the flow curve is applied, stable shear flow cannot be sustained. The sample splits up into coexisting bands supporting different shear rates (or viscosities) and different  microstructures \cite{cates_adv,fielding_SM,olmsted_rheolacta}. This phenomenon is called `shear banding'. The critical value of the shear rate at which shear bands form is  typically the inverse of  the micellar relaxation time. It can be seen that if a reduced stress ${\sigma_{red}}$, denoted by the short, dashed line in Fig. \ref{fig:shear}(A), is applied to the sample, homogeneous flow cannot be sustained and the sample breaks up into bands that support shear rates $\dot\gamma_{1}$ and $\dot\gamma_{2}$, respectively.  Experimentally, this metastability shows up as a stress plateau in the flow curve.  The existence of shear bands was confirmed experimentally with a velocity profiling technique using nuclear magnetic resonance (NMR) \cite{callaghan_gwm}, birefringence measurements, rheology and small angle neutron scattering (SANS) \cite{decruppe_gwm,cappalaere_gwm}. When a constant shear rate is imposed in the metastable region of the flow curve, the sample splits up into bands that form layers along the direction of the flow gradient. This is called gradient banding and has been studied in detail in solutions of GWMs \cite{cates_adv,rehage_gwm,spenley_gwm,berret_gwm}. If a constant shear stress is applied, the sample breaks up into bands with layer normals in the vorticity direction. This phenomenon is known as vorticity banding \cite{goveas_gwm}. When salt is added to the sample, the metastable region of the flow curve develops a  non-zero slope $\alpha$  ($\sigma \sim {\dot\gamma}^{\alpha}$) \cite{ranjini_lang,ganapathy_gwm} due to an enhanced concentration difference between the shear bands \cite{fielding_epje}.

 It was shown in \cite{bandyopadhyay_gwm} that when an aqueous, semi-dilute solution of GWMs is sheared at a constant rate that lies in the metastable region of the flow curve, the time-series of the stress relaxation fluctuates deterministically and is characterised by a finite correlation dimension and a positive Lyapunov exponent. This phenomenon is called  'rheological chaos'.  It was proposed that a modified version of the Johnson-Segalman model \cite{js}, incorporating terms accounting for the coupling between the mean micellar length and the shear rate, the dynamics of the mechanical interfaces and the flow-concentration coupling, could be used to model this phenomenon.  Subsequent experiments have also demostrated that chaotic flows can be supported by semi-dilute solutions of giant wormlike micelles \cite{berret1_gwm,bandyopadhyay_gwm,ganapathy_gwm}, dilute, shear-thickening solutions of cylindrical micelles \cite{bandyopadhyay_thick}, lamellar phases of surfactant solutions \cite{wunenberger_lamellar,salmon_lamellar}, concentrated colloidal suspensions \cite{lootens_denscol}, granular matter \cite{fenistein_band,jaeger_band} and foam \cite{foam_band}.

The effect of a strong coupling between flow and concentration on the development of chaos was investigated in \cite{ganapathy_gwm}. The flow curve of a salt-free GWM solution (an aqueous solution of the cationic surfactant cetyltrimethylammonium tosylate, CTAT)  shows a stress plateau and is represented by solid circles in Fig. \ref{fig:shear}(B).  The flow curve obtained after salt is added to the CTAT solution is represented by hollow symbols in the same plot. In the experiment with salt, the stress varies strongly with shear rate in the metastable region (the flow curve is plotted in the inset of Fig. \ref{fig:shear}(B) shows a slope $\alpha \sim$ 0.32) \cite{ganapathy_gwm,fielding_epje,helfand_flow}. Stress relaxation data, acquired by imposing fixed shear rates in the metastable region of the flow curve,  are displayed in Fig. \ref{fig:shear}(C) and exhibit rapid temporal fluctuations. In simultaneous SALS experiments for VV and VH polarizations, the strong flow-concentration coupling  results in  characteristic butterfly patterns, stretched along the direction of flow \cite{ganapathy_gwm}. Interestingly, the temporal oscillations observed in the intensity data acquired in the VH polarisation experiments exhibit the same trends as those observed for stress fluctuations at the same shear rate. The Fourier analysis of the stress relaxation  data  acquired at $\dot\gamma$ = 23 s$^{-1}$  and 25 s$^{-1}$ (panels (a) and (b) of Fig.  \ref{fig:shear}(C)) show two-frequency quasiperiodicity \cite{ott_chaos}, with further analysis pointing to Type-II intermittency. The stress relaxation data acquired at $\dot\gamma$ = 27 s$^{-1}$ (panel (c) in Fig. \ref{fig:shear}(C))  is chaotic with a positive Lyapunov exponent ($\lambda$ = 0.14) and an exponential Fourier spectrum. For a review on recent efforts in the theoretical modelling of the complex dynamics of shear banded flows, the reader is referred to \cite{fielding_SM} and the references therein.

\subsection {Flow-induced birefringence:}

\hspace{0.25cm}Rheo-optic techniques, like the simultaneous SALS imaging and mechanical rheometry experiments of the micellar samples discussed in section 3.4, are extremely important in the study of soft materials as they connect the microscopic structures of these materials with their macroscopic rheology \cite{macosko_rheo}. Most soft materials have anisotropic chemical structures and therefore anisotropic polarisabilities. When anisotropic molecules orient in a particular direction (by the imposition of a flow, for example) the refractive index $n$ is seen to depend on the polarisation of light. This is called birefringence. When a beam of light containing two orthogonal polarisations is incident on a birefringent material, the birefringence $\Delta{n}$ is defined as $n_{1}$ - $n_{2}$, where $n_{1}$  and $n_{2}$ are the two refractive indices of the material  measured along two orthogonal axes.  

In 1813, D. Brewster observed  birefringence in glasses and gels that were subjected to flow \cite{brewster_FIB}. The stress-optical relation (SOR), which arises from the sensitive dependence of the microscopic structure of soft materials  on the externally imposed flow,  postulates that birefringence varies linearly and isotropically with the applied stress \cite{maxwell_FIB,lodge_fib,peterlin_annrev}. For a shear flow, the SOR is written as $\sigma_{xy} = (1/2C)\Delta{n}\sin(2\chi)$, and $\sigma_{xx} - \sigma_{yy} = (1/C)\Delta{n}\cos(2\chi)$, where $\sigma_{xy}$ and $\sigma_{xx}-\sigma_{yy}$ denote the shear stress and the first normal stress difference respectively, $\chi$ is the extinction angle and $C$ is the stress-optic coefficient \cite{macosko_rheo}. The validity of the SOR was verified by shearing concentrated solutions of polyisoprene \cite{pearson_FIB} and polymeric melts of polyethylene (PE) and PS \cite{kriegel_fib, chai_fib,macosko_rheo}. FIB in a thermotropic polymer undergoing the isotropic-nematic transition was studied in \cite{mather_FIB}. Above a critical elongational flow rate, an aqueous solution of the polyelectrolyte poly(acrylamide-co-sodium acrylate) (PAA) with externally added salt shows a negative birefringence value. PAM solutions, in contrast, show positive birefriengence values \cite{farinato_FIB}. FIB  has been reported in a mixture of lyotropic liquid crystals in the isotropic phase  \cite{fernandes_FIB} and in shear banded GWM solutions \cite{lee_FIB,decruppe_gwm}. In dilute aqueous surfactant solutions with strongly binding counterions, shear-thickening occurs above an induction time $\tau_{i}$. Here, $\tau_{i}$ is essentially the time required to form optically birefringent shear-induced structures (SIS) \cite{wang_fib,wunderlich_fib}. A detailed review of the optical effects generated by flow can be found in \cite{peterlin_annrev}.

\subsection {Soft glassy rheology:}

\hspace{0.25cm} The glass transition in supercooled liquids is characterised by an  increase in the sample viscosity by several orders of magnitude as the temperature is lowered towards the glass transition temperature $T_{g}$ \cite{angell_plot}.  $T_{g}$ is accepted widely to be the temperature at which the viscosity of the supercooled liquid is 10$^{13}$ Poise \cite{IUPAC}. An increase in the volume fraction $\phi$ of a colloidal suspension results in a dramatic enhancement of the sample's viscosity \cite{hunter_glass} and is reminiscent of the glass transition phenomenon observed in supercooled liquids. Increasing the volume fraction of colloidal suspensions, therefore, is equivalent to decreasing the temperature of glass-forming  liquids. It was demonstrated by Marshall and Zukoski that the measured zero shear viscosity $\eta$ of concentrated suspensions of sterically stabilized silica colloids diverges at $\phi_{g} \rightarrow$ 0.638  and obeys the Doolittle equation for glassy flow: $\eta \sim C\exp[D\phi/(\phi_{g}-\phi)]$ \cite{marshall_glass}. 

The rate at which $\eta$ (or $\tau$)  approaches $T_{g}$ defines the fragility of a material \cite{angell_plot,debend_glass}. Most soft glassy materials (SGMs) show non-Arrhenius dependences of $\eta$ and $\tau$ upon $\phi$ and are called fragile glasses \cite{angell_plot,vogel_glass,fulcher_glass,tammann_glass}. SGMS are characterised by spatially correlated dynamics or `dynamic heterogeneity' which is associated with a breakdown of the Stokes Einstein relation between diffusion coefficient and viscosity. Confocal microscopy experiments have established the presence of spatial regions of increased cooperativity that grow as the sample approaches the glass transition \cite{crocker_dh}. This process is accompanied by a dramatic slowing down of  $\tau$. Similar to supercooled and glassy materials, temporal relaxations in SGMs are characterised by two-step response functions \cite{segre_PRL,band_SM,doi_pol,cippelletti_coll}. The short-time dynamics is identified with a `beta' relaxation process and represents the jostling of the particle trapped in a cage formed by its neighbours. The long-term dynamics is identified with the `alpha' relaxation process. The alpha dynamics slows down considerably as the material evolves or `ages' and owes its origin to cooperative rearrangement events.

\begin{figure}[!t]
\begin{center}
\includegraphics[width=1.1\columnwidth]{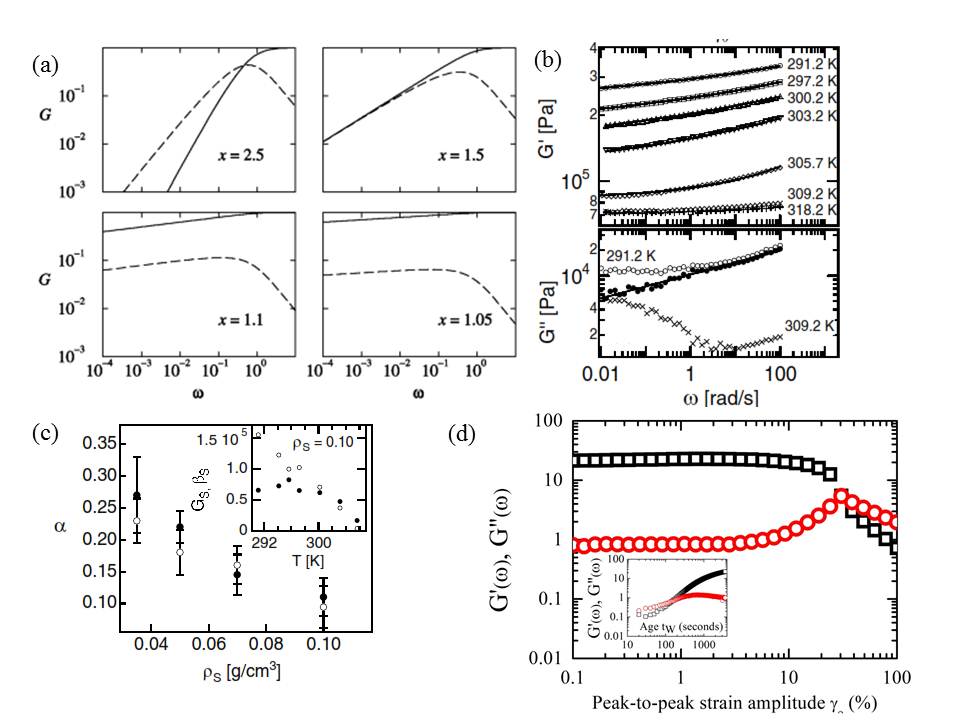} 
\caption{(a): $G^{\prime}$ (solid lines) and $G^{\prime\prime}$ (dashed lines) for different noise temperatures $x$ are reproduced with permission from \cite{sollich_sgr}. (b) $G^{\prime}$ (top panel) and $G^{\prime\prime}$ (bottom panel) of a thermotropic liquid crystal octylcyanobiphenyl (8CB), confined within the pores of an aerosil gel of density $\rho_{s}$ = 0.10 g/cc, are shown for several temperatures. The isotropic to nematic transition temperature T$_{NI}$ of 8CB is 313.98 K and the nematic to smectic transition temperature T$_{NA}$ is 306.97 K. The  $G^{\prime}$ curves acquired for 8CB samples in the smectic phase fit the relation $G^{\prime}(\omega) = G_{g} + G_{s} + \beta_{s}\omega^{\alpha}$ (fits are shown by solid lines). $G_{g}$ and $G_{s}$ are the contributions from the aerosil gel and the 8CB smectic respectively and $\alpha = x$  - 1. The fit of $G^{\prime\prime}(\omega)$ to the SGR prediction after subtraction of the nematic contribution is also shown in the bottom panel of (b). (c) The decrease in $\alpha$ with increase in aerosil concentration $\rho_{s}$. (b) and (c) are adapted from \cite{bandyo_sgr}. (d) $G^{\prime}(\omega)$ (squares) and  $G^{\prime\prime}(\omega)$ (circles), acquired from a  5 wt.\% montmorillonite clay (Bentonite) suspension  \cite{samim_mmt}, display the features predicted in \cite{miyazaki_epl}.}
\label{fig:sgr}
\end{center}
\end{figure}

Foams, emulsions, concentrated colloidal suspensions and slurries are  some examples of SGMs. These materials are all  characterised by structural disorder and metastability \cite{sollich_sgr} and their rheology displays certain peculiar properties. Their elastic and viscous moduli,  $G^{\prime}(\omega)$ and $G^{\prime\prime}(\omega)$ respectively, show weak power-law dependences on the applied angular frequency $\omega$. This behaviour, which persists down to the smallest accessible frequencies, violates linear response theory \cite{kramers_cause,kronig_cause}. As the glass transition is approached, both moduli become flatter. Starting from a trap model \cite{bouchaud_trap} in which the interactions are described in terms of a mean-field `noise temperature' $x$, it was shown in  \cite{sollich_sgr} that  $G^{\prime} \sim \omega^{x-1}$ for $1 < x < 3$, and   $G^{\prime\prime} \sim \omega^{x-1}$ for $1 < x < 2$ in the limit of small  $\omega$. For larger $x$, Boltzmann behaviour is recovered. These features are illustrated in Fig. \ref{fig:sgr}(a). In this model, steady shear behaviour can be described as follows: for $x<$ 1, a non-zero yield stress $\sigma_{\circ}$ appears when $\dot\gamma \rightarrow$ 0, while for $\dot\gamma >>$ 1,   $\sigma - \sigma_{\circ} \sim \dot\gamma^{x-1}$. Power-law fluid behaviour ($\sigma \sim \dot\gamma^{x-1}$) is seen for 1$ <x <$2. Newtonian flow is recovered for $x > 2$. When a thermotropic liquid crystal 8CB is confined to the pores of an aerosil gel, its moduli show power-law variations with frequency \cite{bandyo_sgr}. Increasing the aerosil concentration decreases the noise temperature of the system. These results are displayed in Figs. \ref{fig:sgr}(b) - (c). In another work,  qualitative agreement of the linear rheology data acquired from aging clay suspensions with the predictions of SGR was reported \cite{bonn_sgr}.

Amorphous materials are not in thermodynamic equilibrium below $T_{g}$ \cite{kovacs_glass}. Instead, these materials exhibit an extremely slow approach to equilibrium through a very gradual reduction in their free volumes and mobilities. A consequence of this aging behaviour is a slow evolution of the physical properties of these materials. L. C. Struik \cite{struik_glass} pointed out that aging continues well below $T_{g}$. The mechanical properties of the glass are, therefore, functions of the aging time. Furthermore, it is seen that all polymers age similarly when small strains are applied.  Since mobility is inversely proportional to the relaxation time of a material, creep and stress relaxation curves shift to lower waiting times as the sample ages. A simple horizontal shifting along the waiting time axis can be used to superpose all the curves \cite{struik_glass}. The relaxation time increases as a linear function of $t_{w}$ ($\tau \sim t_{w}^{\mu}$, where $\mu$ = 1) in a process called `simple aging'. Interestingly, the stress relaxation curves of linear polymer melts, acquired at several temperatures, can be shifted horizontally on a  logarithmic time axis to yield a master curve spanning several decades in time. This is called  time-temperature superposition \cite{williams_tts}.

SGMs, being inherently out-of-equilibrium, also display aging. An aqueous foam ages by a combination of three processes: bubble coarsening (which arises from the diffusion of gas from smaller to larger bubbles), bubble collapse (which leads to a reduction of the surface energy of the bubbles) and drainage (the settling of water under gravity) \cite{weaire_foam}. In the coarsening process, the average bubble radius $<R>$ evolves with waiting time $t_{w}$ according to the relation: $<R> \sim t_{w}^{0.5}$ \cite{durian_foamscience}. $t_{w}$ is sometimes also referred to as the idle time of the sample, and is a measure of the time elapsed since sample preparation. When approximately 1-3 \% of Laponite clay is stirred in water, the sample evolves gradually and spontaneously from a liquid-like to a solid-like (soft glassy) consistency \cite{olphen_clay}  due to the gradual evolution of the electrostatic interactions in the system \cite{bandyopadhyay_clay} that causes random, localized, stress relaxation events in the sample \cite{knaebel_clay}.  When aqueous foams coarsen, strains are induced in the system. If the local yield strain is exceeded, the bubbles rearrange with a relaxation time $\tau$. Diffusive wave spectroscopy (DWS) data for aqueous shaving foams shows $\tau \sim t_{w}^{\mu}$, with $\mu \approx$ 1 \cite{durian_foampre}. Although the aging mechanism of clay suspensions is very different from that of aqueous foams, $\tau$ for an aging clay suspension, estimated using multispeckle dynamic light scattering (MDLS), also follows the relaxation $\tau \sim t_{w}^{\mu}$, with $\mu \approx $1 \cite{bellour_pre}. For smaller waiting times, hyper-aging dynamics ($\mu  >$ 1) has been observed \cite{shahin_langmuir}.\\
\begin{figure}[!t]
\begin{center}
\includegraphics[width=0.9\columnwidth]{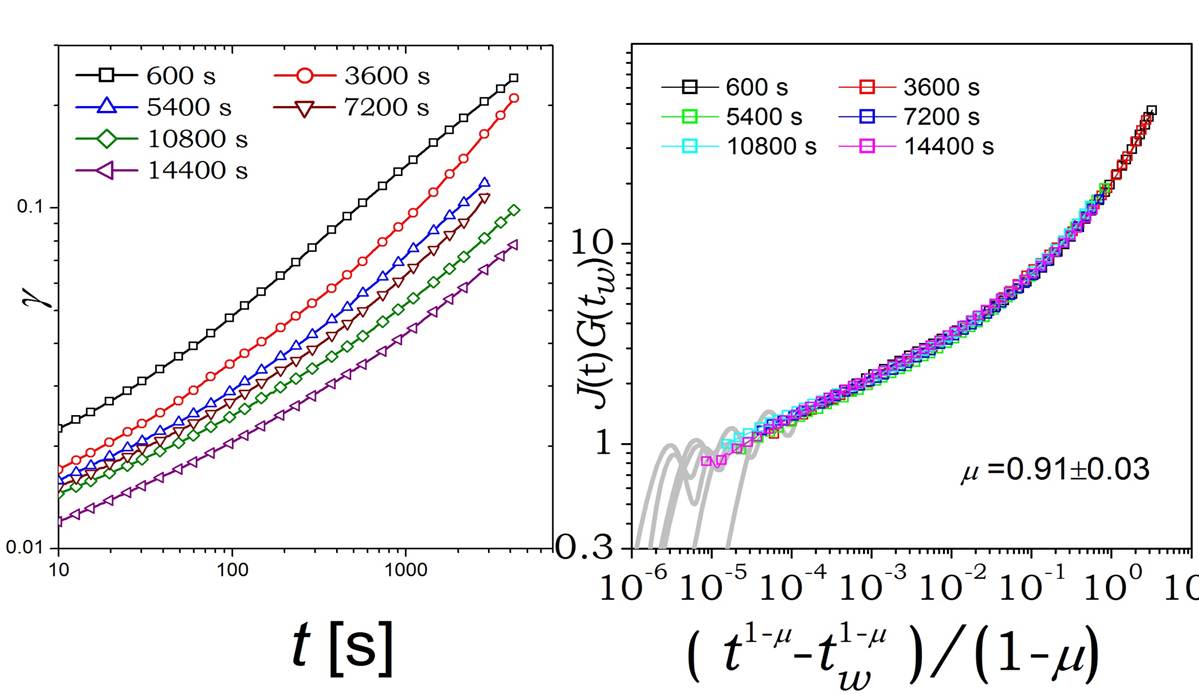} 
\caption{The left panel shows creep data for emulsion  paint at several idle times. The right panel displays the data collapse using the effective time approach \cite{joshi_prl,yogesh}. The value of $\mu$ is very close to 1 and was predicted in \cite{struik_glass}.}
\label{fig:yogesh}
\end{center}
\end{figure}
At low ionic concentrations, clay suspensions form repulsive Wigner glasses comprising randomly oriented clay discs interacting through screened Coulombic interactions. Addition of salt to these suspensions screens the inter-particle electrostatic repulsion and results in the formation of gel networks. Glasses and gels that are subjected to oscillatory shears exhibit distinct rheological signatures \cite{miyazaki_epl}. At the lowest strains, both $G^{\prime}$ and G$^{\prime\prime}$ are independent of the applied oscillatory strain amplitude $\gamma$. As $\gamma$ is increased while keeping $\omega$ fixed, G$^{\prime\prime}$ shows a peak. This is followed by an yielding process that is characterised by power-law decays of $G^{\prime}$ and $G^{\prime\prime}$ and $G^{\prime\prime} > G^{\prime}$ \cite{miyazaki_epl}. These features can be explained using mode coupling theory (MCT) arguments \cite{gotze_mct}. Representative data for a typical SGM is displayed in Fig. \ref{fig:sgr}(d).

 Metastable soft materials  do not obey time translational invariance.  The Boltzmann superposition principle, which states that the strain response of a complex loading is the sum of the strains due to each load, is not obeyed by these systems \cite{macosko_rheo,barnes_rheo}.  Recently, an effective time approach was proposed in \cite{joshi_prl} for the prediction of the rheology of SGMs over short and long time scales. The left panel of Fig. \ref{fig:yogesh} shows the creep data for emulsion paint at several ages, while the right panel shows superposition of this data when scaled  using the effective time approach.  Time-temperature superposition, irreversible aging and idle time-aging time-salt concentration superposition for aging clay suspensions was also reported \cite{yogesh_sm,yogesh_langmuir}. These studies are extremely important as they demonstrate that it is, in fact, possible to successfully predict the slow dynamics of SGMs. 

\subsection{Granular jets, granular streams and the segregation of vibrated granular mixtures}

\hspace{0.25cm} Granular media are the most ubiquitous examples of complex systems that are inherently out of equilibrium \cite{degennes_sand}.  Sand, rice, glass beads and mustard seeds are some examples of this class of systems.   In his book on grains and powders, J. Duran defines a granular medium as a conglomeration of discrete, solid,  macroscopic particles that interact through predominantly dissipative mechanisms such as static friction and inelastic collisions \cite{duran_book}.  As a result, granular media display unusual features that are unexpected in regular solids, liquids and gases \cite{jaeger_rmp}. Unlike in traditional media, not all contacts in granular media carry forces. Instead, forces are transmitted through networks of particle contacts called force chains. As a consequence, the weight of granular materials can be supported by the walls of the container and the rate of flow of grains through an orifice is independent of the pressure head \cite{janssen_forcechains}.  The distribution of contact forces in a granular solid was investigated in \cite{corwin_nature} and the network of force chains was directly visualised using  elasto-birefringent disks in \cite{behringer_nature}.  The velocity and position distribution functions of granular media in the very dilute limit (`granular gases') can display Maxwell Boltzmann statistics \cite{olafsen_natmat}. However, unlike in conventional gases, they exhibit clustering and collapse. The Maxwell's demon has been demonstrated in  \cite{wolf_book}. 

 The gravitational potential energy of a grain in a pile is several orders of magnitude larger than its thermal energy. Granular packings are therefore metastable. The redundancy of thermal energy sets a lower grain size limit of one micron \cite{degennes_sand}. The dynamics are therefore driven not by  entropy, but by the imposition of external forces. Granular packings exhibit a logarithmically slow approach to the steady state upon external forcing \cite{knight_tap}.   In contrast to colloidal suspensions, solvent-mediated interactions are not important in granular systems. The repulsive inter-grain interactions that exist  in dry, non-cohesive granular media are of substantially shorter range than in charge-stabilised colloidal suspensions.  

Granular materials are fragile and cannot support certain incremental loadings without plastic rearrangements  \cite{cates_fragility}.  A heap of sand is stable as long as its slope is less than the angle of repose. If the angle of repose is exceeded, avalanches of sand, only a few grain diameters wide, flow down along the surface of the heap. These systems are excellent paradigms of driven dissipative systems and form patterns when excited. There are numerous  demonstrations of the complex dynamical  behaviour of granular media in the literature. This section discusses three novel phenomena that highlight the liquid-like response of driven granular media. These are the formation of granular jets when a projectile impacts a bed of sand, the breakup of a granular stream into droplets and the size segregation that is seen in vibrated granular mixtures.

\begin{enumerate}[(a)]

\item{\it Granular jets:} When a projectile hits the surface of a  bed of sand, a transient, axisymmetric crater forms. Grains of sand from the sides move in with radial velocities $v_{r} \sim 1/r$ to fill the crater. As the crater closes, a pressure spike is created and a granular jet shoots out of the bed in the vertically upward direction \cite{shen_physfluid}. Granular jets are reminiscent of the Worthington jets that form when rain drops fall on puddles in a light rain \cite{worthington}. A rain drop falling on the puddle creates a crater. The crater fills up immediately with water that flows in from the sides and a jet of water shoots up vertically \cite{brenner_drop}. An important point to note here is that in contrast to rain drops, grains have no surface tension.  Granular jets are narrower than liquid jets with maximum heights that depend upon the grain size. Viscous, gravitational and inertial forces all contribute to their formation. A steel ball dropped on a bed of loosely packed fine sand (grains of size 40 $\mu$m packed at a volume fraction of 41\%)  creates a crown-shaped granular splash and an impact crater \cite{lohse_prl}. As seen in \cite{shen_physfluid}, the collapse of the crater is followed by the formation of a vertical jet that emerges from the centre of the splash. The formation of clusters at the bottom of the jet is attributed to the inelastic inter-grain collisions. The jet subsequently collapses into a granular heap at the point of impact. This is followed by a violent granular eruption that completely erases the granular heap that formed after the collapse of the jet. 

\indent The penetration of a ball in a loosely packed bed of sand was studied in \cite{twente_dryquicksand}. The ball sinks to a depth of many ball diameters in this laboratory version of `dry quicksand'. The penetration depth of the ball scales with its mass and sand jets are seen to emerge only when the mass of the ball is higher than a threshold value. 

\begin{figure}[!t]
\begin{center}
\includegraphics[width=1.1\columnwidth]{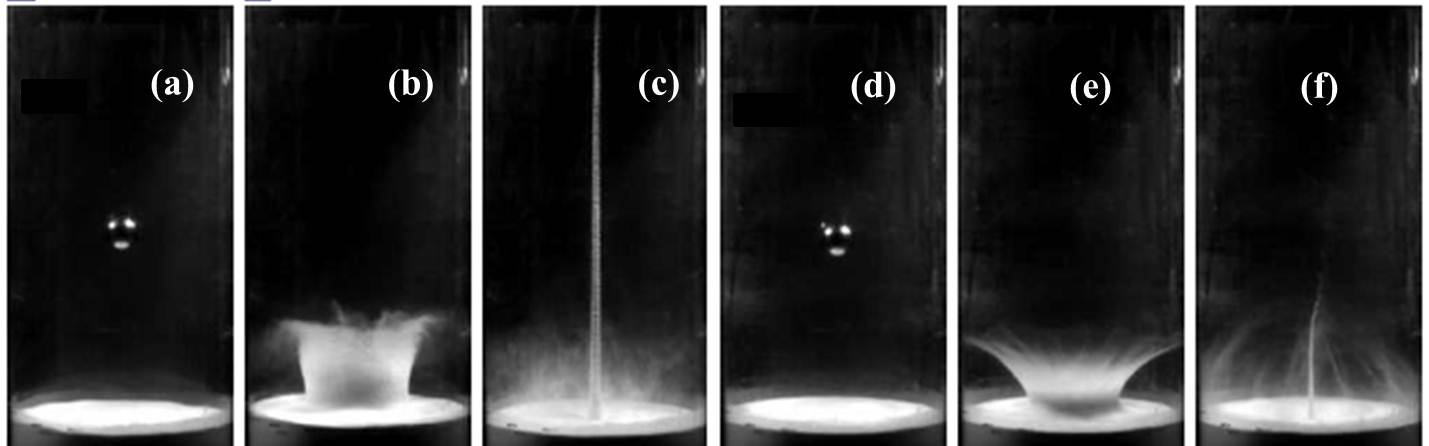} 
\caption{(a)-(c): Images of the formation of granular jets under atmospheric pressure conditions (101 kPa). (a) The projectile drop is followed by (b) an initial splash 0.06 s after the projectile impacts. (c) A  jet that emerges 0.31 s after impact. (d)-(f) are captured under reduced pressure conditions (2.7 kPa). (d) shows the projectile drop, which is followed by (e) a splash whose shape is different from that seen  in (b). (f) The jet that emerges in this experiment is smaller and thinner. The figure is reprinted from \cite{royer_jet} with permission from Macmillan Publishers Ltd.}
\label{fig:jet}
\end{center}
\end{figure}

\indent The importance of the role of interstitial air in the formation of granular jets was established by high speed X-ray radiography and digital video imaging in \cite{royer_jet,royer_epl}. These experiments report the presence of {\it two} jets when a projectile hits a bed of loosely packed sand: a thin, wispy jet whose size does not depend on the air pressure, and an air pressure-dependent thick jet. The thin jet forms from the gravity-driven collapse of the crater. The collapse of this jet is accompanied by the trapping of air bubbles in the sand. These air pockets, compressed by the collapsing grains of sand, drive the upward propulsion of the thick granular jet. Pressure gradients keep the jet collimated, giving rise to an `effective surface tension'.  Fig. \ref{fig:jet} shows the emergence of granular jets following the impact of a projectile on a sand bed under two different air pressure conditions. The complex interplay between interstitial gas, bed particles and the impacting sphere has been studied in detail in \cite{royer_epl,royer_pre}. Videos of granular jets can be viewed at \cite{uchi_url} and a review on sand jets can be found in \cite{jaeger_jet}.\\

\begin{figure}[!t]
\begin{center}
\includegraphics[width=0.8\columnwidth]{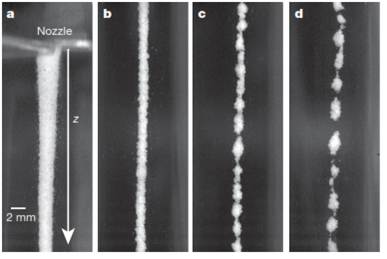} 
\caption{(a) A stream of 107 $\pm$ 19 $\mu$m glass balls dropping through a nozzle of diameter 4 mm. The granular stream, (b) 20 cm, (c) 55 cm and (d) 97 cm below the nozzle, are shown. The formation of droplets and their breakup are clearly visible in (c) and (d). The figure is reprinted from \cite{royer_drop} with permission from Macmillan Publishers Ltd.}
\label{fig:stream}
\end{center}
\end{figure}

\item{\it Granular streams:} A thin stream of Newtonian liquid breaks up into droplets. This phenomenon is called the Rayleigh Plateau instability and owes its origin to molecular surface tension \cite{shi_rayleigh}.   A recent high speed imaging experiment of a thin granular stream emerging from a hopper orifice, peformed  in the co-moving frame, confirms that granular streams can also break up into droplets \cite{royer_drop}. It must be noted here that the liquid-like behaviour of granular media falling under gravity was first reported in 1890 \cite{khamontoff}.

\indent Glass beads of sizes 107 $\pm$ 19 $\mu$m are dropped through a nozzle of diameter 4 mm \cite{royer_drop}.  The granular stream, first stretched by gravity, eventually develops density inhomogeneities. The stream subsequently breaks up into droplets that are connected by narrow bridges. These bridges rupture as the clusters continued to separate. The formation of granular droplets is particularly intriguing, as sand is made up of dry, non-cohesive grains and is not expected to have a surface tension. 

\indent The nature of the granular clusters do not change when the air pressure conditions are changed \cite{mobius_cluster}. Grain-gas interactions are therefore not responsible for droplet formation. The compactness of the clusters decreases when the grain inelasticity is increased. This counterintuitive observation rules out dissipation due to the inelastic collisions of grains as the mechanism driving cluster formation. The inter-grain cohesive forces were controlled by modifying the surface roughness of the grains, the humidity and by using different materials. Atomic force microscopy (AFM) experiments show that the formation of  granular droplets is driven by extremely weak inter-grain cohesive forces that arise due to attractive van der Waals interactions and result in the formation of inter-grain capillary bridges. These cohesive forces give rise to a force similar to surface tension whose magnitude ($\sim$ 10$^{-1} \mu$N/m) is four to five orders of magnitude lower than the surface tension of common liquids. As these cohesive forces are extremely weak, the droplets stretch under gravity and eventually rupture. Unlike in liquids, where the breakup of droplets is driven by thermal fluctuations, granular clustering is driven by collisional cooling. These observations are presented in Fig. \ref{fig:stream} and are extremely important in the understanding of ultra-low surface tension processes \cite{lohse_news}. 

\indent It should be noted here that `granular surface tension' was also proposed in \cite{kellay_prl}. By drawing analogies between the small-scale interface fluctuations of  granular streams flowing under gravity and  thermally induced capillary waves, the authors estimate a granular surface tension of magnitude $\sim$ 100 $\mu$N/m.

\item{\it Granular size segregation:} 
\begin{figure}[!t].
\begin{center}
\includegraphics[width=0.8\columnwidth]{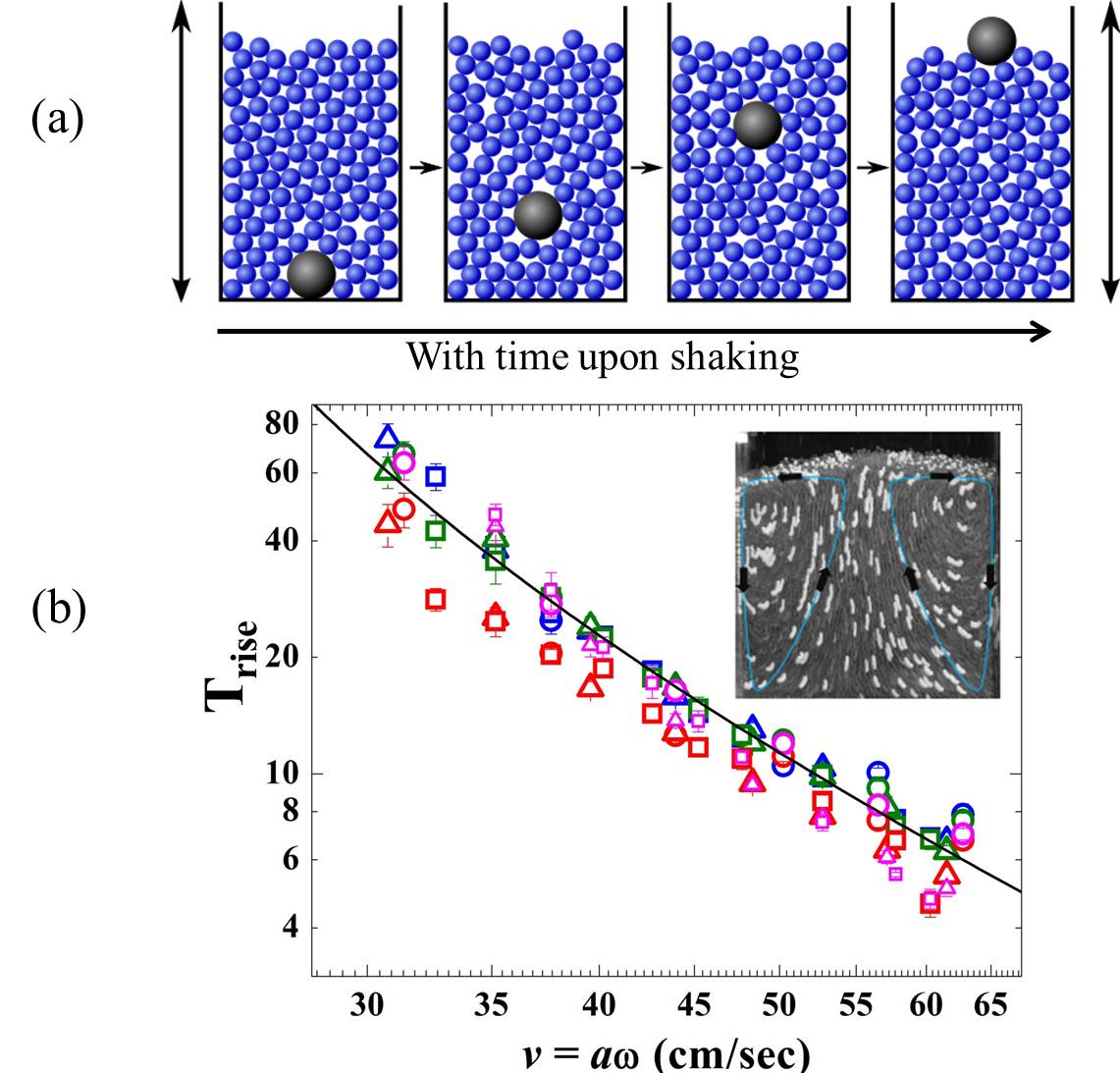}  
\caption{(a) The Brazil nut effect (BNE), reproduced from \cite{wiki_bne} under a Creative Commons License. (b) $T_{rise}$ of an  intruder through a vertically vibrated bed of small seeds scales with the bed velocity. The solid line shows a fit to the form T$_{rise} \sim (v-v_{c})^{\alpha}$, where $v_{c}$ is the shaking velocity of the bed that is required for the onset of bulk convective rolls, and $\alpha$ = 2. Different symbols  denote different angular frequencies $\omega$ and different colours denote intruders of different densities. The inset shows the convection rolls (the lines with arrows are guides to the eye) that are generated in the system by vertical vibrations. This image is obtained by overlaying several consecutive high-speed  images of the vibrated granular bed of small seeds. The white mustard seeds, distributed randomly in a bed of dark mustard seeds, act as tracers. The figure is reproduced from \cite{hejmady_pre}.}
\label{fig:bne}
\end{center}
\end{figure} 
When a bowl of mixed nuts is shaken, the largest nut, the `intruder', always rises to the top. This phenomenon is called the Brazil nut effect (BNE) or the muesli effect (Fig. \ref{fig:bne}(a)). The segregation of granular mixtures by shaking is a phenomenon that  has been historically expoited by mankind.  Foodgrains are often separated from chaff by shaking in the wind. It was also noted a long time ago that when coal was transported in a wagon, the bigger pieces of coal were more likely to come out on top. In this context, it was reported that the act of shaking a granular bed generates voids that are more easily filled up by the smaller bed particles. The larger particle therefore rise to the top  \cite{brown}.  In addition to void-filling mechanisms \cite{williams_voids,rosato}, granular size segregation can be driven by granular convection \cite{ehrichs_science}. The phenomenon is  is also quite sensitive to the presence of interstitial air \cite{mobius_pre}.

 For a comprehensive review of granular size segregation, the reader is referred to \cite{kudrolli_review}.  When a bed of small beads is vertically shaken at a fixed acceleration, three distinct regimes are observed for the rise of an intruder  \cite{vanel_prl}. In the first convective regime, heaping is observed. In this regime, $T_{rise}f \sim e^{f}$, where T$_{rise}$ is the time that the intruder takes to rise to the top of the bed from the bottom (the `rise time') and  $f$ is the frequency of the sinusoidal vibration.  For larger $f$, a second convective regime, where $T_{rise}f \sim e^{f^{2}}$, is observed. As $f$ is increased further, there is a the third regime where $T_{rise}$ depends on the size ratio of the intruder and the bed particles. In this non-convective regime, the upward rise of the intruder is driven by the structural defects that form due to the presence of the intruder. The rise of a large, heavy intruder through a bed of small intruders has been explained as a reverse buoyancy effect in \cite{shinbrot_prl}. In this work, the dominant force driving segregation is void-filling. A large but light intruder, in comparison, exhibits large fluctuations in its motion. This discourages void-filling and the intruder is seen to sink to the bottom of the container. 

\indent Large, heavy intruders, embedded in a bed of smaller particles, can sink to the bottom of the vibrated container. This is the reverse Brazil nut effect (RBNE) \cite{hong_prl}. The masses, densities and inelasticities of the bed and intruder particles,  the shape of the container and the dimensionless acceleration $\Gamma$  ($= a\omega^{2}/g$, where $a$ and $\omega = 2\pi f$ are respectively the amplitude and the angular frequency of the oscillatory vibration and $g$ is the acceleration due to gravity) are major factors that drive BNE and RBNE \cite{breu_prl,shinbrot_prl,mobius_nature}. $T_{rise}$ of an intruder in a vibrated bed of glass beads is reported to be nonomonotonic with intruder density in \cite{mobius_nature} under atmospheric pressure conditions. The nonmonotonicity disappears as the pressure is reduced, thereby highlighting the importance of the role of interstitial air in driving BNE.  In a quasi two-dimensional experiment, it is seen that $T_{rise} \sim (\rho/\rho_{m})^{-1/2}$, where $\rho/\rho_{m}$ is the ratio of the densities of the intruder and the bed particles \cite{liffman_gm}. 

\indent The segregation process can be reversed (BNE can change to RBNE)  if the walls of the container are changed from vertical to inclined (for example, from a rectangular box to a funnel). The roughness of the side-walls is also very important. The friction of the grains with the walls can produce convection rolls that drop down the walls and rise up the centre. An intruder that is larger than the thin, downward granular stream can be carried by the convection roll to the top of the container where it stays trapped, thus producing the BNE \cite{knight_prl}. When a binary granular mixture is vibrated vertically keeping $f$ fixed, but at different $a$ values, the small particles are seen to accumulate at the bottom of the container for low values of $a$ (when 1$<\Gamma<$3.5). For larger $a$ (when $\Gamma >$ 3.5), the small particles rises to the top (RBNE) \cite{breu_prl}. 

\indent For a quasi two-dimensional bed of small particles vertically vibrated at a small $\Gamma$, the rise of the intruder is driven by the successive formation and destruction of granular arches \cite{duran_pre}. For large $\Gamma$, the intruder rise is driven by granular convection. The transition between these two regimes depends on the size ratio of the intruder and the bed particles. High speed imaging for a quasi two-dimensional geometry shows that convective motion can exist at low $\Gamma$ \cite{cooke_pre}. In these experiments, the intruder and the bed particles are seen to move up with the same velocity. The upward movement is facilitated by the existence of slip planes and the block movements of particles slipping past each other. For small $\Gamma$ and size ratios, the rise is intermittent. For large $\Gamma$ and size ratios, the rise is continuous. The effects of drag forces on granular segregation was investigated by inducing gas flow through the perforated base of a granular bed \cite{liu_prl} . More recently,  purely convection-driven segregation was uncovered in the rise of a large intruder through a vibrated granular bed packed with tiny seeds \cite{hejmady_pre}. This work reports that T$_{rise} \sim (v - v_{c})^{-\alpha}$, where $v_{c}$ is the minimum shaking velocity required for the onset of bulk convection rolls in the granular bed, and the constant  $\alpha$ depends on the aspect ratio of the bed. The scaling holds for different sizes and densities of the intruders, for different shapes of the small bed particles and for different side-wall roughnesses. The value of $\alpha$ depends on the aspect ratio of the bed. Fig. \ref{fig:bne} (b) shows that T$_{rise}$ data for many intruder densities can be scaled with the velocity of shaking $v$ as long as $v > v_{c}$. The inset shows the bulk convective rolls that push the intruder to the top of the bed. The intruder therefore behaves like an approximately massless particle that tracks the flow of the convecting seeds. The role of particle-scale rearrangements was investigated in \cite{harrington_arxiv}. It is reported in this work that bidisperse granular mixtures in a three-dimensional geometry segregate under steady, but not under cyclical, shear.

\indent The segregation of granular mixtures subjected to swirling horizontal excitations was reported in \cite{aumaitre_pre}. Increased humidity is not conducive to granular size segregation \cite{williams_voids,liffman_gm}. The addition of a small amount of liquid to a granular medium results in the formation of bridges at the granular contacts \cite{schiffer_nature} and completely destroys BNE \cite{kudrolli_wet}. Viscous and capillary forces are very important in the segregation of wet granular media. When the grains are immersed in a liquid of low viscosity, a transition back to segregation is observed.

\end{enumerate}

\subsection{Pattern formation}

Patterns such as arrays, cracks, fractals, spirals and ripples are common in nature and are characterised by a regularity in form  \cite{wiki_pattern,fineberg_pattern}. Mathematicians, physicists and biologists have always taken a keen interest in the emergence of patterns. Joseph Plateau studied the topology and geometry of soap films using the concept of minimal surfaces \cite{weaire_foam} and Alan Turing worked extensively on Fibonacci phyllotaxis, morphogenesis \cite{turing_morpho} and oscillatory chemical reactions (such as the Belousov Zhabotinsky reaction \cite{belousov,zhabotinsky}). For a detailed review of the formation of spatiotemporal patterns in hydrodynamic systems (such as in Taylor-Couette and Rayleigh- Benard flows), parametric-wave instabilities, excitable biological media {\it etc.}, the reader is referred to \cite{cross_rmp}. 
 
Granular media, excited by vertical and horizontal vibrations, can exhibit a wealth of patterns \cite{kruelle_review}. Bulk convection rolls in vibrated granular media have been demonstrated in \cite{ehrichs_science}. Convection, heaping and cracking have been observed experimentally in vertically vibrated granular slurries \cite{smith_slurry}. Surface waves travelling against gravity have been reported in  noncohesive granular media when the Couette cell containing the grains is vibrated at large $a$ and $\Gamma$  \cite{behringer_prl}. Vertically oscillated granular layers in an evacuated container show a sequence of patterns. As $\Gamma$ is increased at a fixed $f$,  squares, hexagons, kinks and ultimately a disordered state are observed \cite{melo_prl}. Two-dimensional localized states or oscillons, which are excitations that are driven by collisions between the highly dissipative grains of sand, are observed in a vibrating layer of sand \cite{umban_nature}. These oscillons, which can assemble into molecular or crystalline structures, coexist with a pattern-free state  and oscillate at a frequency $f$/2 between conical peaks and craters. Oscillons of the same phases repel, while those of different phases attract.  They can associate to form dipoles, chains and lattices. Using a two-dimensional Swift-Hohenberg formalism, it was shown in \cite{craw_oscillon} that oscillons can develop and interact in any medium that has reflection and discrete time-translational symmetry, and that can undergo large transition hysteresis \cite{liou_pattern,hunt_review}. 

Both oscillons and dissipative solitary states (DSS) have been observed in vertically vibrated clay suspensions \cite{liou_pattern}. DSS are seen in highly dissipative fluids and are large-amplitude, highly localised propagating states with the same periodicity as that of the driving. Blowing compressed air on the surface of these suspensions, while increasing $\Gamma$ at a fixed $f$, produces a hysteretic transition from a featureless to an oscillon state. When $f$ is increased, the oscillons interact to form complex localised structures like doublets and triad patterns \cite{liou_pattern}.   An increase in  $\Gamma$ results  in the formation of  stripes. Unlike in granular media, the oscillons and stripe patterns that develop in vibrated colloidal suspensions exist over the entire range of frequencies explored.  A further increase in $\Gamma$, however, leads to the destabilisation of the stripes and the formation of large finger-like protrusions. If $\Gamma$ is now decreased, by decreasing $f$ below a threshold value at which the amplitude of the oscillon approaches the suspension depth $h$, DSS is observed. In vibrated cornstarch suspensions, persistent holes appear at high $f$  when a finite perturbation is applied  \cite{merkt_cornstarch}. When $f$ is increased, the rim of the hole destabilises to form fingerlike protrusions. A further increase in $\Gamma$ results in the  delocalisation of the holes. The whole surface of the cornstarch suspension is then covered by erratic undulations. It was shown in \cite{deegan_pre} that stress hysteresis contributes to the formation of persistent holes in particulate systems. 

Several studies of the formation of patterns at the morphologically unstable interface between two fluids in a porous medium or in a Hele Shaw geometry have been reported \cite{wiki_finger,saffmann_finger,hele_shaw}.   Coloured water injected into aqueous colloidal clay suspensions form fractal viscous fingers (VF) for low clay concentrations \cite{damme_nature,lemaire_vef}. A transition to a viscoelastic fracturing (VEF) regime is seen when clay concentration is increased \cite{lemaire_vef}.  The VEF patterns reported in \cite{lemaire_vef} have branching angles of 90$^{\circ}$ with the main crack, and are characterised by fractal dimensions that are lower than those of VF patterns. The fingers that form when a less viscous liquid percolates a viscoelastic polymer solution are much narrower \cite{lindner1}. Viscous fingers with highly branched morphologies form when less viscous air displaces foams and colloidal gels in a radial Hele Shaw geometry at very low rates of flow \cite{lindner2}. 

Polydisperse glass beads, of sizes 50-100 $\mu$m and suspended in water-glycerol mixtures, are filled in a radial Hele-Shaw cell at different filling fractions $\phi$ \cite{sandnes_prl}. When the liquid mixture is withdrawn using a syringe pump, air invades the grain-liquid mixture and labyrinthine patterns form. The characteristic  length scales of the patterns decrease with $\phi$ and increase with the gap widths of the Hele-Shaw geomtery. Numerical simulations are performed to verify that the competition between frictional and capillary forces drives the formation of the observed patterns. 

\begin{figure}[!t].
\begin{center}
\includegraphics[width=1\columnwidth]{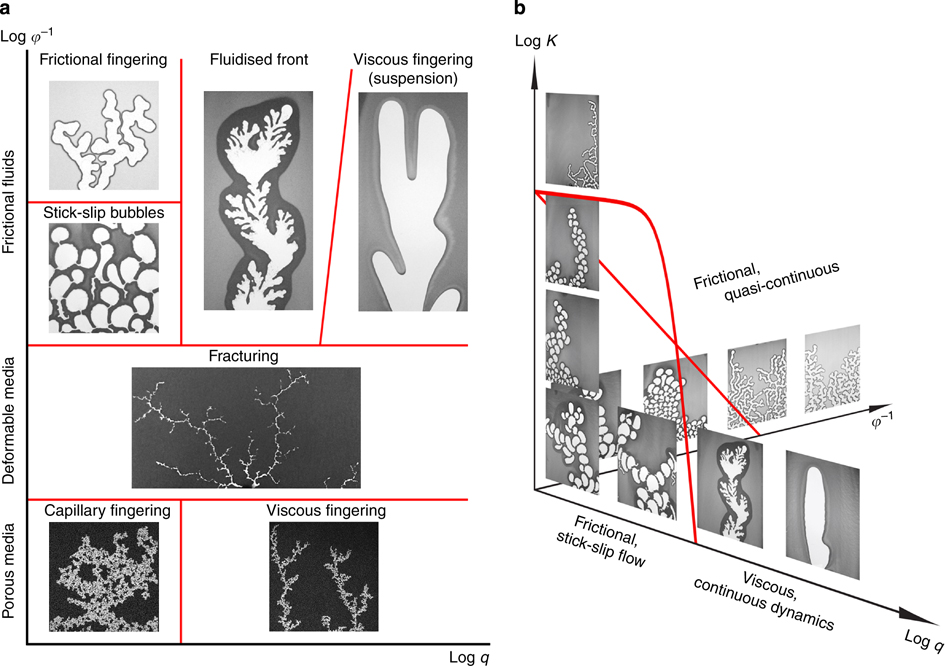}  
\caption{(a) A phase diagram in the $\phi^{-1}-q$ plane shows several contrasting patterns. (b) A pictorial representation of the displacement dynamics of frictional fluids in the $\phi^{-1}-q-K$ plane. The figure is reproduced from \cite{sandnes_frictionalflow} under a Creative Commons license.}
\label{fig:granpatt}
\end{center}
\end{figure}

It has been mentioned earlier that the surface tension of granular matter is negligible when compared to ordinary liquids. This requires the simultaneous existence of a local cusp structure and a global fractal structure in granular fingering patterns \cite{wiegmann}. This was confirmed experimenally for a radial Hele Shaw flow in \cite{nagel_granularpatt}.  The width of the fingers formed at the interface between two Newtonian liquids were reported to decrease with the local interface velocity $v$ as $v^{-1/2}$. The width of granular fingers, in contrast, grow as $v^{1/2}$  \cite{nagel_granularpatt} and the fractal dimensions of the patterns lie between 2 and 1.7 (the diffusion limited aggregation or DLA value). When air invades a bed of water-saturated glass beads, three different invasion regimes, driven by capillary fingering, viscous fingering and capillary fracturing, are observed. These regimes depend upon the injection rate of air, the size of the glass beads and the confining stress \cite{juanes_prl}. A review on the collective behaviour of grains and the formation of patterns can be found in \cite{aronson_rmp}. In another experiment, when air invades a frictional granular bed in which glass beads mixed with a viscous liquid are allowed to settle, the interfacial dynamics shows a range of behaviours from intermittent to quasi-continuous to continuous \cite{sandnes_frictionalflow}. In this work, capillary fingering, viscous fingering and fracturing regimes are observed as the flow rate $q$ of the injected air, the filling fraction $\phi$ and stiffness $K$ of the glass beads  are changed.  The results are displayed in Fig. \ref{fig:granpatt}.

The formation of patterns by highly dense assays of active filaments driven by hydrodynamic forces was reported in \cite{schaller_pattern}.  The application of a strong electric field resulted in the formation of patterns at polymer-air-polymer thin film interfaces \cite{pol_patt}. The formation of desiccation cracks in clay-polymer films attached to a substrate was studied in \cite{tarafdar_cracks}.

\section{Conclusions}

This review presents an introductory discussion of a selective list of novel phenomena that have been observed in soft materials. The nonintuitive dynamics of soft systems is illustrated by highlighting several experiments, many of which, such as the demonstrations of shear-thinning and thickening, polymer rod climbing, die swell and the tubeless siphoning effects, can  be performed with commonly available soft materials such as cake batter, egg yolk and cornstarch suspensions.  This is followed by many more examples of novel time-dependent phenomena that are exhibited by soft matter and are reported in the literature. These include chaotic stress fluctuations in sheared wormlike micellar solutions, flow-induced birefringence, the existence soft glassy rheology, and the formation of patterns. Given the space constraints, the discussions are, by no means, complete. A comprehensive list of journal and review articles that describe each topic in much more detail has been included for further reference wherever possible.

\section{Acknowledgements}

The author is extremely grateful to Yogesh Joshi for his critical comments regarding this article, for sharing his raw data on emulsion paint (presented here as Fig.10), and for extending all possible help in performing experiments to demonstrate the Weissenberg effect, the Barus effect and Fano flow (these images are displayed in Fig. 4 and the videos are uploaded as supporting information). The author thanks Samim Ali, Manish Kaushal, Prakhyat Hejmady, Rajib Basak and Debasish Saha for their help at various stages during the preparation of this manuscript.

\section{Bibliography}

%\begin{verbatim}

%\end{verbatim}
% 


\begin{thebibliography}{99}


\bibitem{hooke_elasticity}
R. Hooke, in {\it Lectures de Potentia Restitution} (John Martyn, London, 1678).

\bibitem{newton_viscosity}
I. S. Newton, in {\it Philosphiae Naturalis Principia Mathematica} (Royal Society of London, 1687).


\bibitem{markovitz_rheo} H. Markovitz, {\it Physics Today}, {\bf 21}, 23 (1968).


\bibitem{macosko_rheo}
C. W. Macosko, in {\it Rheology- Principles, Measurements and Applications} (Wiley-VCH, New York, 1994).

\bibitem{barnes_rheo}
H. A. Barnes, J. F. Hutton and K. Walters, in {\it An Introduction to Rheology} (Elsevier, USA, 1989).

\bibitem{mezger_rheo}
T. Mezger, in {\it The Rheology Handbook} (Vincentz, Hannover, 2006). 

\bibitem{dorai_rheo} D. Doraiswamy, {\it Rheol. Bull.}, {\bf 71}, 1 (2002).

\bibitem{piazza_soft} R. Piazza, in {\it Soft Matter- The Stuff that Dreams are made of} (Springer-Verlag, Italy, 2010).

\bibitem{mitov_soft} M. Mitov, in {\it Sensitive Matter- Foams, Gels, Liquid Crystals and Other Miracles} translated by Giselle Weiss, (Harvard University Press, 2012).

\bibitem{hamley_soft} I. W. Hamley, in {\it Introduction to Soft Matter: Synthetic and Biological Self-Assembling Materials} (John Wiley, 2007).

\bibitem{degennes_nobel} P.-G. de Gennes, {\it http://www.nobelprize.org/nobel{$_{-}$}prizes/physics/laureates/1991/gennes{$-$}lecture.pdf}, 1991.

\bibitem{goodyear_patent}
C. Goodyear, in {\it Improvement in India-Rubber Fabrics} (Patent no. 3633, United States Patent Office, 1844).

\bibitem{degennes_fragile} P.-G de Gennes, in {\it Fragile Objects- Soft Matter, Hard Science and the Thrill of Discovery}, (Springer-Verlag, New York, 1994).

\bibitem{rajib} R. Basak and R. Bandyopadhyay (unpublished).

\bibitem{ncfm_rheologynon} https://www.youtube.com/watch$?$v=Ol6bBB3zuGc\&feature=related. This is a National Council of Fluid Mechanics film.

\bibitem{landau_elasticity} L. D. Landau and E. M. Lifshitz, in {\it Theory of Elasticity} (Butterworth-Heinemann, Oxford, 1986).

\bibitem{feinman_vol2} R. D. Feynman, R. B. Leighton and M. Sands, in {\it Lectures on Physics} (Narosa Publishing House, New Delhi, 1986) p. 1195. 

\bibitem{weber_silk} W. Weber, {\it Ann. Phys. Chem.}, {\bf 34}, 247 (1835).

\bibitem{pol_sillputt} The originator of this figure, adapted from http://www.me.gatech.edu/jonathan.colton/me4210/sillyputty.pdf, is Cambridge Polymer Group Inc., Boston, MA, USA. 

\bibitem{voigt_linear} M. A. Meyers and K. K. Chawla, in {\it Mechanical Behaviour of Materials} (Prentice Hall, London, 1999) p. 570.

\bibitem{maxwell_linear} J. C. Maxwell, {\it Phil. Trans. Roy. Soc. Lond.}, {\bf 157}, 49, (1867).

\bibitem{laurent_petri} L. Courbin, E. Denieul and H. A. Stone, {\it J. Stat. Mech}, N10001 (2006).





\bibitem{youtube_cornstarch1} http://www.youtube.com/watch?v=GorX5iVxHAw\&feature=related.

\bibitem{fall_prl} A. Fall, N. Huang, F. Bertrand, G. Ovarlez and D Bonn,  {\it Phys. Rev. Lett.}, {\bf 100}, 018301 (2008).  

\bibitem{liu_jam} A. J. Liu and S. R. Nagel, {\it Nature}, {\bf 396}, 21 (1998).

\bibitem{jaeger_nature} S. R. Waitukaitis and H. M. Jaeger, {\it Nature}, {\bf 487}, 205 (2012).

\bibitem{gopal_prl} A. D. Gopal and D. J. Durian, {\it Phys. Rev. Lett.}, {\bf 91}, 188303 (2003).

\bibitem{band_SM} R. Bandyopadhyay, P. H. Mohan and Y. M. Joshi, {\it Soft Matter}, {\bf 6}, 1462  (2010).

\bibitem{reynolds_dilatancy} O. Reynolds, {\it Phil. Mag. Series}, {\bf 20}, 469 (1885).

\bibitem{crayola} Silly Putty$^{TM}$ is a registered trademark of Crayola LLC.

\bibitem{szevgai} F. Schalek, A. Szegvari, {\it Kolloid Z}, {\bf  33}, 326 (1923).

\bibitem{schalek_thixo} W. H. Bauer and E. A. Collins, in {\it Rheology: Theory and Applications} (Academic Press, New York, 1967) ch. 8.

\bibitem{rheol_cambridge} J. Mewis and N. J. Wagner, in {\it Colloidal Suspension Rheology} (Cambridge University Press, England, 2012).

\bibitem{barnes_thixo} H. A. Barnes, {\it J. Non-Newtonian Fluid. Mech}, {\bf 70}, 1 (1997).

\bibitem{ferry_polymer} J. D. Ferry, in {\it Viscoelastic Properties of Polymers} (J. Wiley and Sons, New York, 1970).

\bibitem{cox_thin} H. W. Cox and C. W. Macosko, {\it AIChE J.}, {\bf 20}, 785 (1974).

\bibitem{weitz_foam} D. A. Weitz, {\it Nature}, {\bf 381}, 475 (1996). 

\bibitem{kaye_thin}A. Kaye, {\it Nature}, {\bf 197}, 1001 (1963).

\bibitem{versluis_kaye} M. Versluis, C. Blom, D. van der Meer, K. Van der Weele and D. Lohse D,  {\it J. Stat. Mech.},  P07007 (2006).

\bibitem{versluis_youtube} http://www.youtube.com/watch$?$v=GX4$_{-}$3cV$_{-}$3M9w.

\bibitem{bingham_yield} E. C. Bingham, {\it U.S. Bureau of Standards Bull.}, {\bf 13}, 309 (1916). 

\bibitem{goyon_yield} J. Goyon, A. Colin, G. Ovarlez, A. Ajdari and L. Boquet, {\it Nature}, {\bf 453}, 84 (2008).

\bibitem{boquet_yield} L. Boquet, A. Colin and A. Ajdari, {\it Phys. Rev. Lett.}, {\bf 103}, 036001 (2009).

\bibitem{fall_yield} A. Fall, J. Parades and D. Bonn, {\it Phys. Rev. Lett.}, {\bf 105}, 225502 (2010).

%\bibitem{wiki_bingham} http://en.wikipedia.org/wiki/Bingham$_{-}$plastic.
 
\bibitem{coussot_yield} P. Coussot, Q. D. Nguyen, H. T. Huynh, and Daniel Bonn, {\it Phys. Rev. Lett.}, {\bf  88}, 175501 (2002).

\bibitem{chao_pol} K. K. Chao, C. A. Child, E. A. Grens and M. C. Williams, {\it Am. Inst. Chem. Engg. Jl.}, {\bf 30}, 111 (1984).

\bibitem{bird_rheo} R. B. Bird, R. C. Armstrong and O. Hassager, in {\it Dynamics of Polymeric Liquids} (Wiley, New York, 1987).

\bibitem{trouton_extvis} F. T. Trouton, {\it Proc. Royal Soc. A}, {\bf 77}, 426 (1906).


\bibitem{ptoday_bird} R. B. Bird and C. F. Curtiss, {\it Phys. Today}, {\bf 37}, 36 (1984).

\bibitem{khan_rheo} S. A. Khan, J. R. Royer and S. R. Raghavan, in {\it Aviation Fuels with Improved Fire Safety: A Proceedings} (The National Academy of Sciences, USA, 1997), p. 31.

\bibitem{mit_demo} http://web.mit.edu/nnf/research/phenomena/Demos.pdf. 

\bibitem{yogesh_lab} The experiments were performed by Mr. S. Ali with samples prepared by Mr. M. Kaushal in the group of Dr. Y. M. Joshi, Department of Chemical Engineering, Indian Institute of Technology, Kanpur, India. 

\bibitem{yogesh_videos} The .mpg videos that have been uploaded are entitled Supporting video of Weissenberg effect (10.7 MB), Supporting video of Barus effect (12 MB) and Supporting video of Fano flow (8.8 MB).

\bibitem{Weissenberg} K. Weissenberg, {\it Nature}, {\bf 159}, 310 (1947).

\bibitem{kundu_weissenberg} P. K. Kundu, {\it J. Rheol.}, {\bf 17}, 343 (1973). 

\bibitem{mit_weissenberg} http://web.mit.edu/nnf/research/phenomena/rod$_{-}$climb$_{-}$highres.jpg.

\bibitem{tanner_dieswell} R. I. Tanner, in {\it Engineering Rheology} (Clarendon, Oxford, 2000).

\bibitem{youtube_barus} http://www.youtube.com/watch$?$v=iFSF1WXoTG4.

\bibitem{cloitre_JNNFM} M. Cloitre, T. Hall, C. Mata and D. D. Joseph, {\it J. Non-Newtonian Fluid Mech.}, {\bf 79}, 157 (1998).


\bibitem{fano_ext} G. Fano, {\it Arch. Fisiol.}, {\bf 5}, 365 (1908).

\bibitem{tumblr_fano} http://www.tumblr.com/tagged/fano$-$flow.

\bibitem{toms_drared} B. Toms, Proc. {\it Intl. Rheol. Cong.}, {\bf 2}, 135 (1948).

\bibitem{gyr_drared} A Gyr and H. W. Bewersdoff, in {\it Drag Reduction of Turbulent Flow by Additives} (Dordrecht, Kluwer, 1995).

\bibitem{bark_jfm} F. H. Bark, E. J. Hinch and M. T. Landahl, {\it J. Fluid Mech.}, {\bf 68}, 129 (1975).

\bibitem{fas_dragred} http://www.fas.org/irp/agency/dod/jason/dragred.pdf.

\bibitem{bon2_drared} O. Cadot, D. Bonn and S. Douady, {\it Phys. Fluids}, {\bf 10}, 426 (1998).

\bibitem{frings_drared} B. Frings, {\it Rheol. Acta}, {\bf 27}, 92 (1988).

\bibitem{lumley_review} J. L. Lumley, {\it J. Polymer Sci: Macromol. Rev}, {\bf 7}, 263 (1973).


\bibitem{bonn_dragred} D. Bonn, Y. Amarouchene, C. Wagner, S. Douady and O. Cadot, {\it J. Phys. Condens. Matt.}, {\bf 17}, S1195 (2005).



\bibitem{wagner_epl} C. Wagner, Y. Amarouchene, P. Doyle and D. Bonn, {\it Europhys. Lett.}, {\bf 64}, 823 (2003).

\bibitem{hoff_drared} D. Ohlendorf, W. Inerthal and H. Hoffmann, {\it Rheol. Acta},  {\bf 25}, 468 (1986).

\bibitem{drappier_drared}  J. Drappier, T. Divoux, Y. Amarouchene, F. Bertrand, S. Rodts, O. Cadot, J. Meunier and D Bonn, {\it Europhys. Lett.}, {\bf 74}, 362 (2006).

\bibitem{white_drag} A. White, {\it Nature}, {\bf 214}, 585 (1967).

\bibitem{reynolds_turb} O. Reynolds, {\it Phil. Trans. R. Soc. Lond.}, {\bf  174}, 935 (1883).

\bibitem{groisman_turb} A. Groisman and V. Steinberg, {\it Nature}, {\bf 405}, 53 (2000).

\bibitem{larson_news}  R. G. Larson, {\it Nature}, {\bf 405}, 27 (2000).

\bibitem{landau} D. L. Landau and E. M. Lifshitz, {\it Fluid Mechanics}, (Pergamon, Oxford, 1987).

\bibitem{kalika_jrheol} D. S. Kalika and M. M. Denn, {\it J. Rheol.}, {\bf 31}, 815 (1987).

\bibitem{larson} S. J. Muller, R. G. Larson and E. S. G. Shaqfeh, {\it Rheol. Acta}, {\bf 28}, 499 (1989).

\bibitem{larson_review} R. G. Larson, {\it Rheol. Acta}, {\bf 31}, 213 (1992).

\bibitem{cladis_lc} P. E. Cladis and W. van Sarloos, in {\it Solitons in Liquid Crystals} (Springer, New York, 1992) p. 111.

\bibitem{manneville_nematic} P. Manneville, {\it Mol. Cryst. Liq. Cryst.}, {\bf 70}, 223 (1981). 

\bibitem{fardin_turb} M. A. Fardin, D. Lopez, J. Croso, G. Gregoire, O. Cardoso, G. H. McKinley and S. Lerouge, {\it Phys. Rev. Lett.}, {\bf 104}, 178303 (2010). 

\bibitem{fardin_turb2} M. A. Fardin, B. Lasne, O. Cardoso, G. Gregoire, M. Argentina, J. P. Decruppe, and S. Lerouge, {\it Phys. Rev. Lett.}. {\bf 103}, 028302 (2009). 

\bibitem{groisman_turb2} A. Groisman and V. Steinberg, {\it New J. Phys.}, {\bf 6}, 29 (2004).

\bibitem{groisman_nature2} A. Groisman and V. Steinberg, {\it Nature}, {\bf 410}, 905 (2001).

\bibitem{spenley_gwm} N. A. Spenley, M. E. Cates and T. C. B. McLeish, {\it Phys. Rev. Lett.}, {\bf 71}, 939 (1993); N. A. Spenley, X. F. Yuan and M. E. Cates, {\it J. Phys II}, {\bf 6}, 551 (1996).

\bibitem{ganapathy_gwm} R. Ganapathy and A. K. Sood, {\it Phys. Rev. Lett.}, {\bf 96}, 108301 (2006); R. Ganapathy and A. K. Sood, {\it Langmuir}, {\bf 22}, 11016 (2006).

\bibitem{surfactants} R. Zana, in {\it Dynamics of Surfactant Self-Assemblies: Micelles, Microemulsions, Vesicles and Lyotropic Phases} (CRC Press, USA, 2005).

\bibitem{cates_adv} M. E. Cates and S. Fielding, {\it Adv. Phys.}, {\bf 55}, 799 (2006).


\bibitem{fielding_SM} S. Fielding, {\it Soft Matter}, {\bf 3}, 1262 (2007).

\bibitem{olmsted_rheolacta} P. D. Olmsted, {\it Rhel. Acta}, {\bf 47}, 283 (2008).


\bibitem{callaghan_gwm} P. T. Callghan, M. E. Cates, C. J. Rofe and J. A. F. Smeulders, {\it J. Phys. II}, {\bf 6}, 375 (1996). 

\bibitem{decruppe_gwm} J. P. Decruppe, R, Cressely, R. Makhloufi and E. Cappelaere, {\it Coll. Pol. Sci}, {\bf 273}, 346 (1995); J. P. Decruppe, O. Greffier, S. Manneville and S. Lerouge, {\it Phys. Rev. E}, {\bf 73}, 061509 (2006). 

\bibitem{cappalaere_gwm} E. Cappalaere, J. F. Berret, J. P. Decruppe, R. Cresseley and P. Lindner, {\it Phys. Rev. E}, {\bf 56}, 1869 (1997).

\bibitem{rehage_gwm} H. Rehage and H. Hoffmann, {\it Mol. Phys.}, {\bf 74}, 933 (1991).

\bibitem{berret_gwm} J- F. Berret and G. Porte, {\it Phys. Rev. E}, {\bf 60}, 4268 (1999); J.-F. Berret, in {\it Molecular Gels}, edited by R. G. Weiss, P. Terech (Springer Dordrecht, 2005) p 235.

\bibitem{goveas_gwm} J. L. Goveas and P. D. Olmsted, {\it Eur. Phys. J. E}, {\bf 6}, 79 (2001).



\bibitem{ranjini_lang} R. Bandyopadhyay and A. K. Sood, {\it Langmuir}, {\bf 19}, 3121 (2003).

\bibitem{fielding_epje} S. M. Fielding and P. D. Olmsted, {\it Eur. Phys. J. E}, {\bf 11}, 65 (2003).





\bibitem{bandyopadhyay_gwm} R. Bandyopadhyay, G. Basappa and A. K. Sood, {\it Phys. Rev. Lett.}, {\bf 84}, 2022 (2000).

\bibitem{js} M. Johnson and D. Segalman, {\it J. Non-Neton. Fluid Mech.}, {\bf 2}, 255 (1977).

\bibitem{berret1_gwm} J.-F. Berret, {\it Langmuir}, {\bf 13}, 2227 (1997).

\bibitem{bandyopadhyay_thick} R. Bandyopadhyay and A. K. Sood, {\it Europhys. Lett.}, {\bf  56}, 447 (2001).

\bibitem{wunenberger_lamellar} A. S. Wunenberger, A. Colin, J. Leng, A. Arneodo and D. Roux, {\it Phys. Rev. Lett.}, {\bf 86}, 1374 (2001).

\bibitem{salmon_lamellar} J.-B. Salmon, A. Colin and D. Roux, {\it Phys. Rev. E}, {\bf 66}, 031505 (2002).

\bibitem{lootens_denscol} D. Lootens, H. Van Damme and P. Hebraud, {\it Phys. Rev. Lett.}, {\bf 90}, 178301 (2003).

\bibitem{fenistein_band} D. Fenistein and M. van Hecke {\it Nature (London)}, {\bf 425}, 256 (2003).

\bibitem{jaeger_band} D. Mueth, G. Debregeas, G. Karczmar, P. Eng, S. R. Nagel, and H. M. Jaeger, {\it Nature}, {\bf 406}, 385 (2000);  X. Cheng, J. B. Lechman, A. F. Barbero, G. S. Grest, H. M. Jaeger, G. S. Karczmar, M. E. Mobius, and S. R. Nagel, {\it  Phys. Rev. Lett.}, {\bf 96}, 038001 (2006).

\bibitem{foam_band} K. Krishna and M. Dennin, {\it Phys. Rev. E}, {\bf 78}, 051504 (2008).

\bibitem{helfand_flow} E. Helfand and G. H. Fredrickson, {\it Phys. Rev. Lett.}, {\bf 62}, 2468 (1989).

\bibitem{ott_chaos} E. Ott, in {\it Chaos in Dynamical Systems} (Cambridge University Press, England, 1993).

\bibitem{brewster_FIB} D. Brewster, {\it Phil. Trans. Roy. Soc.}, {\bf 103}, 101 (1813).

\bibitem{maxwell_FIB} J. C. Maxwell, {\it Trans. Roy. Soc.}, {\bf 20}, 87 (1853).

\bibitem{lodge_fib} A. S. Lodge, {\it Nature}, {\bf 176}, 838 (1955).

\bibitem{peterlin_annrev} A. Peterlin, {\it Ann. Rev. Fluid Mech.}, {\bf 8}, 35 (1976). 

\bibitem{pearson_FIB} D. S. Pearson, A. D. Kiss, and L. J. Fetters, {\it J. Rheol.}, {\bf 33}, 517 (1989);   D. S. Pearson, A. D. Kiss, and L. J. Fetters, {\it J. Rheol.}, {\bf 34}, 613 (1989).

\bibitem{kriegel_fib} H. Janeschitz-Kriegel, in {\it Polymer Melt Rheology and Flow Birefringence}, (Springer Verlag, Berlin, 1983).

\bibitem{chai_fib} C. K. Chai, J. Creissel and H. Randrianantoandro, {\it Polymer}, {\bf 40}, 4431 (1999).

\bibitem{mather_FIB} P. T. Mather, H. G. Jeon, C. D. Han and S. Chung, {\it Macromol.}, {\bf 35}, 1326 (2002).

\bibitem{farinato_FIB} R. S. Farinato, {\it Polymer}, {\bf 29}, 2122 (1988).

\bibitem{fernandes_FIB} P. R. G. Fernandes and A. M. Figueiredo Neto, {\it Phys. Rev. E}, {\bf 51}, 567 (1995).

\bibitem{lee_FIB} J. Y. Lee, X.-F. Yuan, G. Fuller and N. E. Hudson, {\it J. Rheol}, {\bf 89}, 537 (2005).

\bibitem{wang_fib} Y. Hu, S. Q. Wang and A. M. Jamieson, {\it J. Rheol.}, {\bf 37}, 531 (1993).

\bibitem{wunderlich_fib} I. Wunderlich, H. Hoffmann and H. Rehage, {\it Rheol. Acta}, {\bf 26}, 532 (1987).

\bibitem{angell_plot} C. A. Angell, {\it Science}, {\bf 267}, 1924 (1995).

\bibitem{IUPAC} IUPAC (International Union of Pure and Applied Chemistry) Compendium of Chemical Terminology, {\bf 66}, 583 (1984)

\bibitem{hunter_glass} G. L. Hunter and E. R. Weeks, {\it Rep. Prog. Phys.}, {\bf 75}, 066501 (2012).

\bibitem{marshall_glass} L. Marshall and C. F. Zukoski, {\it J. Phys. Chem.}, {\bf 94}, 1164 (1990).

\bibitem{debend_glass} P. G. Debendetti and F. H, Stillinger, {\it Nature}, {\bf 410}, 259 (2001).

\bibitem{vogel_glass} H. Vogel, {\it Z. Phys.}, {\bf 22}, 645 (1921).

\bibitem{fulcher_glass} J. Fulcher, {\it J. Am. Ceram Soc.}, {\bf 8},  339 (1925).

\bibitem{tammann_glass} G. Tammann and G. Hess, {\it Z. Anorg. Allg. Chemie}, {\bf 156}, 245 (1926).

\bibitem{crocker_dh} E. R. Weeks, J. C. Crocker, A. C. Levitt, A. Schofield and D. A. Weitz, {\it Science}, {\bf 287}, 627 (2000).

\bibitem{segre_PRL} P. N. Segre, V. Prasad, A. B. Schofield, and D. A. Weitz, {\it Phys. Rev. Lett.}, {\bf 68}, 6042 (2001).

%\bibitem{ranjini_SM} R. Bandyopadhyay, P. H. Mohan and Y. M. Joshi, {\it Soft Matter}, {\bf 6}, 1462 (2010)


\bibitem{doi_pol} M. Doi and S. F. Edwards, in {\it The Theory of Polymer Dynamics} (Oxford University Press, Oxford, 1987).

\bibitem{cippelletti_coll} L. Cippelletti and L. Ramos, {\it J. Phys. Condens. Matt.}, {\bf 17}, R253 (2005).

\bibitem{sollich_sgr} P. Sollich, F. Lequeux, P. Hebraud and M. E. Cates, {\it Phys. Rev. Lett.}, {\bf 78}, 2020 (1997).

\bibitem{kramers_cause} H.A. Kramers, {\it Atti Cong. Intern. Fisica, (Transactions of Volta Centenary Congress)}, {\bf 2}, 545 (1927).

\bibitem{kronig_cause}R. de L. Kronig, {\it J. Opt. Soc. Am.}, {\bf 2}, 545 (1927).

\bibitem{bouchaud_trap} J. P. Bouchaud, {\it J. Phys I (France)}, {\bf 2}, 1705 (1992).

\bibitem{bandyo_sgr} R. Bandyopadhyay, D. Liang, R. H.Colby, J. L. Harden and R. L. Leheny, {\it Phys. Rev. Lett.}, {\bf 94}, 107801 (2005).

\bibitem{bonn_sgr} D. Bonn, D. Ross, S. Hachem, S. Gridel and J. Meunier, {\it Europhys. Lett.}, {\bf 59}, 786 (2002). 

\bibitem{kovacs_glass} A. J. Kovacs, {\it Fortschr. Hochpolym, Forsch.}, {\bf 3}, 394 (1964).

\bibitem{struik_glass} L. C. E. Struik, in {\it Physical Aging in Amorphous Polymers and Other Materials} (Elsevier, Amsterdam, 1978). 

\bibitem{williams_tts} M.L. Williams, R.F. Landel and J.D. Ferry, {\it J. Amer. Chem. Soc.}, {\bf 77}, 3701 (1955).

\bibitem{weaire_foam} D. Weaire and S. Hutzler {\it The Physics of Foam}, (Oxford University Press, USA, 2001).

\bibitem{durian_foamscience} D. J. Durian, D. A. Weitz and D. J. Pine, {\it Science}, {\bf 252}, 686 (1991).

\bibitem{olphen_clay} H. van Olphen, in {\it An Introduction to Clay Colloid Chemistry} (Interscience Publications, New York, 1963).

\bibitem{bandyopadhyay_clay} R. Bandyopadhyay, D. Liang, H. Yardimci, D. A. Sessoms, M. A. Borthwick, S. G. J. Mochrie, J. L. Harden, and R. L. Leheny, {\it Phys. Rev. Lett.}, {\bf 93}, 228302 (2004).

\bibitem{knaebel_clay} A. Knaebel, M. Bellour, J.-P. Munch, V. Viasnoff, F. Lequeux and J. L. Harden, {\it Europhys. Lett.}, {\bf 52}, 73 (2000).


\bibitem{samim_mmt} S. Ali and R. Bandyopadhyay (unpublished).

\bibitem{miyazaki_epl} K. Miyazaki, H. M. Weiss, D. A. Weitz and D. R. Reichman, {\it Europhys. Lett.}, {\bf 75}, 915 (2006).

\bibitem{gotze_mct} W. Gotze and L. Sjorgen, {\it Rep. Prog. Phys.}, {\bf 55}, 241 (1992).


\bibitem{durian_foampre} D. J. Durian, D. A. Weitz and D. J. Pine, {\it Phys. Rev. A}, {\bf 44}, R7902 (1991).

\bibitem{bellour_pre} M. Bellour, A. Knaebel, J. L. Harden, F. Lequeux, and J.-P. Munch, {\it Phys. Rev. E}, {\bf 67}, 031405 (2003).

\bibitem{shahin_langmuir} A. Shahin and Y. M. Joshi, {\it Langmuir}, {\bf 28}, 5826 (2012). 

%\bibitem{poon_gelcol} L. Starrs, W. C. K. Poon, D. J. Hibberd and M. M. Robbins, {\it J. Phys. Condens. Matter}, {\bf 14}, 2485 %(2002). 


\bibitem{joshi_prl} A. Shahin and Y. M. Joshi, {\it Phys. Rev. Lett.}, {\bf 106}, 038302 (2011).

\bibitem{yogesh} Data from \cite{joshi_prl}, obtained {\it via} private communications with Yogesh Joshi.


\bibitem{yogesh_sm} R. Gupta, B. Baldewa and Y. M. Joshi, {\it Soft Matter}, {\bf 8}, 4171 (2012).

\bibitem{yogesh_langmuir} A. Shahin and Y. M. Joshi, {\it Langmuir}, {\bf 26}, 4219 (2010).


\bibitem{degennes_sand}P. G. de Gennes, {\it  Review of Modern Physics}, {\bf 71}, 374 (1999).


\bibitem{duran_book} J. Duran, in {\it Sands, Powders, and Grains: An Introduction to the Physics of Granular Materials}, translated by A. Reisinger  (Springer-Verlag New York, Inc., New York,1999). 

\bibitem{jaeger_rmp} H. M. Jaeger, S. R. Nagel and R. P. Behringer, {\it Rev. Mod. Phys.}, {\bf 68}, 1259 (1996).



\bibitem{janssen_forcechains} H. A. Janssen, {\it Zeitschr. d. Vereines deutscher Ingenieure}, {\bf 39}, 1045 (1895). 

\bibitem{corwin_nature} E. I. Corwin, H. M. Jaeger and S. R. Nagel, {\it Nature}, {\bf 435}, 1075 (2005). 

\bibitem{behringer_nature} T. S. Majmudar and R. P. Behringer, {\it Nature}, {\bf 435}, 1079 (2005).


\bibitem{olafsen_natmat} G. W. Baxter and J. S. Olafsen, {\it Nat. Mater.}, {\bf 425}, 680 (2003). 

\bibitem{wolf_book} H. Hinrichsen and D. E. Wolf, in {\it The Physics of Granular Media} (WileyVCH Verlag GmbH and Co, 2004).


\bibitem{knight_tap} E. Ben-Naim, J. B. Knight, E. R. Nowak, H. M. Jaeger and S. R. Nagel, {\it Physica D}, {\bf 123}, 380 (1998).


\bibitem{cates_fragility} M. E. Cates, J. P. Wittmer, J.-P. Bouchaud, and P. Claudin, {\it Phys. Rev. Lett.}, {\bf 81}, 1841 (1998).



\bibitem{shen_physfluid} S. T. Thorodssen and A. Q. Shen, {\it Phys. Fluids}, {\bf 13}, 4 (2001).

\bibitem{worthington} A. M. Worthington, in {\it A Study of Splashes} (Longmans, Green and Co., London, 1908).

\bibitem{brenner_drop} M. P. Brenner, {\it Nature (London)}, {\bf 403}, 377 (2000).


\bibitem{lohse_prl} D. Lohse, R. Bergmann, R. Mikkelson, C. Zeilstra, D. van der Meer, M. Versluis, K. van der Weele, M. van der Hoef and H. Kuipers, {\it Phys. Rev. Lett.}, {\bf 93}, 198003 (2004).



\bibitem{twente_dryquicksand}  D. Lohse,  R. Rauhe, R. Bergmann and D. van der Meer, {\it Nature}, {\bf 
432}, 689 (2004)

\bibitem{royer_jet} J. R. Royer, E. I. Corwin, A. Flior, M.-L. Cordero, M. Rivers, P. Eng, and H. M. Jaeger, {\it Nat. Phys.} {\bf 1}, 164 (2005). 


\bibitem{royer_epl} J. R. Royer, B. Conyers, E. I. Corwin, P. J. Eng and H. M. Jaeger, {\it  Europhys. Lett.}, {\bf  93}, 28008 (2011).


\bibitem{royer_pre} J. Royer, E. I. Corwin, B. Conyers, A. Flior, M. L. Rivers, P. J. Eng and H. M. Jaeger, {\it Phys. Rev. E}, {\bf 78}, 011305 (2008).

\bibitem{uchi_url} http://jfi.uchicago.edu/granular/jets.html.

\bibitem{jaeger_jet} H. M. Jaeger, {\it Physics World}, {\bf  18}, 34 (2005). 

\bibitem{shi_rayleigh} P.-G. de Gennes, F. Brochard-Wyart and D. Quere, in {\it Capillarity and Wetting Phenomena: Drops, 
Bubbles, Pearls, Waves}  (Springer, 2003); X. Shi, M. P. Brenner and S. R. Nagel, {\it Science}, {\bf 265}, 219 (1994).

\bibitem{royer_drop} J. R. Royer, D. J. Evans, L. Oyarte, Q. Guo, E. Kapit, M. E. M$\ddot{o}$bius, S. R. Waitukaitis, and H. M. Jaeger, {\it Nature}, {\bf 459}, 1110 (2009).

\bibitem{khamontoff} N. Khamontoff, {\it J. Russ. Phys.-Chem. Soc}, {\bf 22}, 281 (1890).

\bibitem{mobius_cluster} M. E. Mobius, {\it Phys. Rev. E}, {\bf 74}, 051304 (2006).

\bibitem{lohse_news} D. Lohse and D. van der Meer, {\it Nature}, {\bf 459}, 1064 (2009).

\bibitem{kellay_prl} Y. Amarouchene, J.-F. Boudet and H. Kellay, {\it Phys. Rev. Lett.}, {\bf 100}, 218001 (2008).



\bibitem{brown} R. L. Brown, {\it J. Inst. Fuel}, {\bf 13}, 15 (1939).

\bibitem{williams_voids} J. C. Williams, {\it J. Fuel Soc.}, {\bf 14}, 29 (1963).

\bibitem{rosato} A. Rosato, K. J. Stranburg, F. Prinz and R. H. Swenden, {\it Phys. Rev. Lett.}, {\bf 58}, 1038 (1987).


\bibitem{ehrichs_science} E. E. Ehrichs, H. M. Jaeger, G. S. Karczmar, J. B. Knight, V. Y. Kuperman, S. R. Nagel, {\it Science}, {\bf 
567}, 1632 (1995).


\bibitem{mobius_pre} M. E.  Mobius, X. Cheng, P. Eshuis, G. S. Karczmar, S. R. Nagel, and H. M. Jaeger, {\it  Phys. Rev.  E}, {\bf 72}, 011304  (2005).

\bibitem{wiki_bne} https://commons.wikimedia.org/wiki/File:Brazil$_{-}$nut$_{-}$effect.svg.


\bibitem{hejmady_pre} P. Hejmady, R. Bandyopadhyay, S. Sabhapandit and A. Dhar, {\it Phys. Rev. E Rapid Comm.}, {\bf 86}, 050301 (2012).


\bibitem{kudrolli_review} A. Kudrolli, {\it Rev. Mod. Phys.}, {\bf 67}, 209 (2004).

\bibitem{vanel_prl} L. Vanel, A. D. Rosato and R. Dave, {\it Phys. Rev. Lett.}, {\bf 78}, 1255 (1997).

\bibitem{shinbrot_prl} T. Shinbrot and F. J. Muzzio, {\it Phys. Rev. Lett.}, {\bf 81}, 4365 (1998).


\bibitem{hong_prl} D. C. Hong, P. V. Quinn and S. Luding, {\it Phys. Rev. Lett.}, {\bf 86}, 3423 (2001).

\bibitem{breu_prl} A. P. J. Breu, H.-M. Ensner, C. A. Cruelle and I Rehberg, {\it Phys. Rev. Lett.}, {\bf 90}, 014302 (2003).

\bibitem{mobius_nature} M. E. Mobius, B. E. Lauderdale and H. M. Jaeger, {\it Nature}, {\bf 414}, 270 (2001).


\bibitem{liffman_gm} K. Liffman, K. Muniandy, M. Rhodes, D. Gutteridge and G. Metcalfe, {\it Granular Matter}, {\bf 3}, 205 (2001).

\bibitem{knight_prl} J. B. Knight, H. M. Jaeger and S. R. Nagel, {\it Phys. Rev. Lett.}, {\bf 70}, 3728 (1993). 


\bibitem{duran_pre} J. Duran, T. Mazozi, E. Clement and J. Rajchenbach, {\it Phys. Rev. E}, {\bf 50}, 5138 (1994).

\bibitem{cooke_pre} W. Cooke, S. Warr, J. M. Huntley and R. C. Ball, {\it Phys. Rev. E}, {\bf 53}, 2812 (1996).

\bibitem{liu_prl} C. Liu, L. Wang, P. Wu and M. Jia, {\it Phys. Rev. Lett.}, {\bf 104}, 188001 (2010).


\bibitem{harrington_arxiv} M. Harrington, J. H. Weijs and W. Losert, aeXiv: 1302.3788 [cond-mat.soft] (2013).

\bibitem{aumaitre_pre} S. Aumaitre, C. A. Cruelle and J. Rehberg, {\it Phys. Rev. E}, {\bf 64}, 041305 (2001).

\bibitem{schiffer_nature} P. Schiffer, {\it Nature Physics}, {\bf 1}, 21 (2005). 

\bibitem{kudrolli_wet} A. Samadani and A. Kudrolli, {\it Phys. Rev. Lett.}, {\bf 85}, 5102 (2000).
 
\bibitem{wiki_pattern} http://en.wikipedia.org/wiki/Patterns$_{-}$in$_{-}$nature.

\bibitem{fineberg_pattern} J. Fineberg, {\it Nature}, {\bf 382}, 763 (1996).

\bibitem{weaire_foam} D. L. Weaire and S. Hutzler, in {\it Physics of Foams}, (Oxford University Press, USA, 2001).

\bibitem{turing_morpho} A. M. Turing, {\it Phil. Trans. Roy. Soc. Lond.}, {\bf 237}, 37 (1952).

\bibitem{belousov} B. P. Belousov, {\it Collection of Abstracts on Radiation Medicine}, {\bf 147}, 145 (1959).

\bibitem{zhabotinsky} A. M. Zhabotinsky, {\it Biophysics}, {\bf 9}, 306 (1964).

\bibitem{cross_rmp} M. C. Cross and P. C. Hohenberg, {\it Rev. Mod. Phys.}, {\bf 65}, 851 (1993).

\bibitem{kruelle_review} C. A. Kruelle, {\it Rev. Adv. Mater. Sci.}, {\bf 20}, 113 (2009).

\bibitem{smith_slurry} J, M. Schleier-Smith and H. A. Stone, {\it Phys. Rev. Lett.}, {\bf 86}, 3016 (2001).


\bibitem{behringer_prl} H. K. Pak and R. P. Behringer, {\it Phys. Rev. Lett.}, {\bf 71}, 1832 (1993).

\bibitem{melo_prl} F. Melo, P. B. Umbanhowar and H. L. Swinney, {\it Phys. Rev. Lett.}, {\bf 75}, 3838 (1995).

\bibitem {umban_nature} P. Umbanhower, F. Melo and H. L. Swinney, {\it Nature (London)}, {\bf 382}, 793 (1996).

\bibitem{craw_oscillon} C. Crawford and H. Riecke, {\it Physica D}, {\bf 129},  83 (1999).

\bibitem{hunt_review} http://guava.physics.uiuc.edu/$~$nigel/courses/569/Essays$_{-}$Fall2008/files/hunt.pdf.

\bibitem{liou_pattern} O. Lioubashevski, Y. Hamiel, A. Agnon, Z. Reches and J. Fineberg, {\it Phys. Rev. Lett.}, {\bf 83}, 3190 (1999).

\bibitem{merkt_cornstarch} F. S. Merkt, R. D. Deegan, D. I. Goldman, E. C. Rericha and H. L. Swinney, {\it Phys. Rev. Lett.}, {\bf 92}, 184501, (2006).

\bibitem{deegan_pre} R. P. Deegan, {\it Phys. Rev. E}, {\bf 81}, 036319 (2010).

\bibitem{wiki_finger} http://en.wikipedia.org/wiki/Viscous$_{-}$fingering.

\bibitem{saffmann_finger} P. G. Saffmann and G. I. Taylor, {\it Proc. Soc. London A}, {\bf 245}, 312 (1958).

\bibitem{hele_shaw} H. J. S. Hele Shaw, {\it Nature (London)}, {\bf 58}, 34 (1898).

\bibitem{damme_nature} H. van Damme, F. Obrecht, P. Levitz, L. Gatineau and C. Laroche, {\it Nature}, {\bf 320}, 731 (1986).

\bibitem{lemaire_vef} E. Lemaire, P. Levitz, G. Daccord, and H. Van Damme, {\it Phys. Rev. Lett.}, {\bf 67}, 2009 (1991). 

\bibitem{lindner1} A. Lindner, D. Bonn, E. Corvera Poire, M. Ben Amar and J. Meunier, {\it J. Fluid Mech.}, {\bf 469}, 237 (2002).

\bibitem{lindner2}  A. Lindner, P. Coussot and D. Bonn, {\it Phys. Rev. Lett.}, {\bf  85}, 314 (2000). 

\bibitem{sandnes_prl}B. Sandnes, H. A. Knudsen, K. J. Maloy and E. G. Flekkoy, {\it Phys. Rev. Lett.}, {\bf 99}, 038001 (2007).

\bibitem{wiegmann} E. Bettelheim, O. Agam, A. Zabrodin and P. Wiegmann, {\it Phys. Rev. Lett.}, {\bf 95}, 244504 (2005).

\bibitem{nagel_granularpatt} X. Cheng, L. Xu, A. Patterson, H. M. Jaeger and S. R. Nagel, {\it Nat. Phys.}, {\bf 4}, 234 (2008). 

\bibitem{juanes_prl} R. Holtzman, M. L. Szulczewski and R. Juanes, {\it Phys. Rev. Lett.}, {\bf 108}, 264504 (2012).

\bibitem{aronson_rmp} I S. Aronson and L. S. Tsimring, {\it Rev. Mod. Phys.}, {\bf  78}, 641 (2006).

\bibitem{sandnes_frictionalflow} B. Sandnes, E. G. Flekkoy,  H. A. Knudsen, K. J. Maloy and H. See, {\it Nat. Comm.}, DOI: 10.1038/ncomms1289 (2011).

\bibitem{schaller_pattern} V. Schaller ,  C. Weber ,  E. Frey and A. R. Bausch, {\it Soft Matter}, {\bf 7}, 3213 (2011).

\bibitem{pol_patt} G. Amarandei,  P. Beltrame,  I. Clancy,  C. O'Dwyer,  A. Arshak,  U. Steiner,  D. Corcoran and U. Thiele, {\it Soft Matter}, {\bf 8}, 6333 (2012).  

\bibitem{tarafdar_cracks}  S. Nag, S. Sinha, S. Sadhukhan, T. Dutta and S. Tarafdar, {\it J. Phys. Condens. Matt.}, {\bf 22}, 015402 (2010).


\end{thebibliography}
\end{document}